\documentstyle[psfig]{mnnd}

\newcommand{\be}{\begin{equation}}
\newcommand{\ee}{\end{equation}}
\newcommand{\ba}{\begin{array}}
\newcommand{\ea}{\end{array}}
\newcommand{\bea}{\begin{eqnarray}}
\newcommand{\eea}{\end{eqnarray}}

\newcommand{\mfrac}[2]{\mbox{$\frac{#1}{#2}$}}

\newcommand{\mbf}[1]{{\bf #1}}
\newcommand{\mcal}[1]{{\cal #1}}
\newcommand{\DS}{\displaystyle}

\begin{document}

\title[A timing formula for main-sequence star binary pulsars]{A timing 
formula for main-sequence star binary pulsars}

\author[N. Wex]{Norbert Wex\footnotemark[1]\\
Max-Planck-Institut f\"ur Radioastronomie, Auf dem H\"ugel 69, 
D-53121 Bonn, Germany}

\maketitle

\begin{abstract} 
In binary radio pulsars with a main-sequence star companion, the spin-induced
quadrupole moment of the companion gives rise to a precession of the binary
orbit. As a first approximation one can model the secular evolution caused by
this classical spin-orbit coupling by linear-in-time changes of the longitude
of periastron and the projected semi-major axis of the pulsar orbit. This
simple representation of the precession of the orbit neglects two important
aspects of the orbital dynamics of a binary pulsar with an oblate companion.
First, the quasiperiodic effects along the orbit, due to the anisotropic
$1/r^3$ nature of the quadrupole potential. Secondly, the long-term secular
evolution of the binary orbit which leads to an evolution of the longitude of
periastron and the projected semi-major axis which is non-linear in time.

In this paper a simple timing formula for binary radio pulsars with a
main-sequence star companion is presented which models the short-term secular
and most of the short-term periodic effects caused by the classical spin-orbit
coupling. I also give extensions of the timing formula which account for
long-term secular changes in the binary pulsar motion. It is shown that the
short-term periodic effects are important for the timing observations of the
binary pulsar PSR B1259--63. The long-term secular effects are likely to
become important in the next few years of timing observations of the binary
pulsar PSR J0045--7319. They could help to restrict or even determine the
moments of inertia of the companion star and thus probe its internal
structure.

Finally, I reinvestigate the spin-orbit precession of the binary pulsar PSR
J0045--7319 since the analysis given in the literature is based on an
incorrect expression for the precession of the longitude of periastron. A 
lower limit of $20\degr$ for the inclination of the B star with respect
to the orbital plane is derived.
\end{abstract}


\footnotetext[1]{Email: wex@mpifr-bonn.mpg.de}


\section{Introduction}

Timing observations of pulsars, i.e.\ the measurement of the time-of-arrival
(TOA) of pulsar signals at a radio telescope, is one of the few high-precision
experiments in astronomy and therefore has a wide-ranging field of interesting
applications (Bell 1996).  The first evidence for the existence of
gravitational waves as predicted by Einstein's theory of gravity (Taylor \&
Weisberg 1989) and the first discovery of extrasolar planets (Wolszczan \&
Frail 1992) are just the most striking examples for the achievements of
high-precision pulsar-timing observations. Approximately 10\% of the known
pulsars are members of binary star system, i.e.\ in orbit around a white
dwarf, neutron star or main-sequence star companion. Timing observations of
these binary pulsars is a powerful tool to study various physical and
astrophysical effects related to binary star motion and stellar evolution. To
extract the maximum possible information from pulsar timing observations, one
needs an appropriate model (timing formula) for transforming each measured
topocentric TOA, $t_{obs}$, to the corresponding time of emission, $T$,
measured in the reference frame of the pulsar. Various timing formulae,
particularly for relativistic binary pulsars, have been developed to describe
radio-pulsar timing observations.

With the discovery of PSR B1259--63 during a high frequency survey of the
Galactic plane by Johnston et al.\ (1992) the first radio pulsar with a
massive, non-degenerate companion was found. PSR B1259--63 is a 48-ms pulsar
in a highly eccentric orbit with the main-sequence Be star SS 2883.  The
second known radio pulsar with a massive, non-degenerate companion is PSR
J0045--7319, discovered in a systematic search of the Magellanic Clouds for
radio pulsars (McConnell et al.\ 1991, Kaspi et al.\ 1994). Some of the
parameters of these two main-sequence star binary pulsars are listed in
Table~1.  For both binary systems significant deviations from a Keplerian
orbit have been detected which are most easily explained by classical
spin-orbit coupling (Lai et al.\ 1995, Kaspi et al.\ 1996, Wex et al.\
1997). Due to their high proper rotation the main-sequence star companions of
PSRs B1259--63 and J0045--7319 show rotational deformation and thus give rise
to a gravitational quadrupole field. As a result of this a coupling between
the orbital angular momentum and the spin of the companion takes place and
leads to a precession of the binary orbit.

In this paper I will shown that the present timing formulae represent only
crude approximations to the orbital dynamics caused by the classical
spin-orbit coupling and that there is a need for a new timing formula to model
the timing observations. Before I focus on the construction of a new
timing formula for main-sequence star binary pulsars a short presentation of
the most important timing models is given.

\begin{table}
\begin{tabular}{lcccc} \hline\hline
                  & \qquad & B1259--63 & \qquad & J0045--7319 \\ 
\hline
$P_b$ (days)      && 1237      && 51.17       \\
$x$ (sec)         && 1296      && 174.3       \\
$e$               && 0.870     && 0.808       \\
$\omega$ (deg)    && 138.7     && 115.3       \\
$i$ (deg)         && 36 or 144 && 44 or 136   \\ 
$m_*$ ($M_\odot$) && $\sim 10$ && $8.8\pm1.8$ \\
$R_*$ ($R_\odot$) && $\sim  6$ && $6.4\pm0.7$ \\ \hline
\end{tabular}
\caption{Parameters of the main-sequence star radio binary pulsars (Johnston 
et al.\ 1994, Kaspi et al.\ 1994, Bell et al.\ 1995). $P_b$, $x$, $e$,
$\omega$, $i$, $m_*$ and $R_*$ denote respectively the orbital period, the
projected semi-major axis, the eccentricity of the orbit, the longitude of
periastron, the inclination of the orbit with respect to the line-of-sight,
the mass of the companion main-sequence star and its radius.}
\end{table}

In a simple spin-down law the pulsar proper time is related to the phase,
${\cal N}$, of the pulsar by
\be
   {\cal N}(T) = {\cal N}_0+\nu T+\mfrac{1}{2}\dot\nu T^2
                 +\mfrac{1}{6}\ddot\nu T^3\,,
\ee
where $\nu$, $\dot\nu$, and $\ddot\nu$ are the rotation frequency of the
pulsar, its first and second time derivative, respectively (spin parameters).

For a single pulsar the timing formula includes terms related to the position
($\alpha,\delta$), proper motion ($\mu_\alpha,\mu_\delta$) and parallax
($\pi$) of the pulsar. Moreover, it contains terms related to relativistic
time dilation and light propagation effects in the solar system and also
corrects for propagation effects in the interstellar medium (Backer 1989,
Taylor 1989, Doroshenko \& Kopeikin 1990):
\be\label{sptf}\ba{l}
   T=t_{obs}+\Delta_C-D/f_b^2 \\[2mm]
   \qquad\quad +\Delta_{R\odot}(\alpha,\delta,\mu_\alpha, 
   \mu_\delta,\pi)+\Delta_{E\odot}+\Delta_{S\odot}(\alpha,\delta) \,.
\ea\ee
$\Delta_C$ corrects for the offset between the observatory clock and the
`Terrestrial Dynamical Time' represented by the best terrestrial standard of
time. $D/f_b^2$ corrects for the dispersive delay in the interstellar medium
at the (barycentric) frequency $f_b$ where $D$ is proportional to the column
density of free electrons between the pulsar and the observer.
$\Delta_{R\odot}$ describes the so called Roemer delay, a change in the time
of flight of the radio signal caused by the motion of the observer in the
solar system reference frame. $\Delta_{E\odot}$ represents the transformation
between `Terrestrial Dynamical Time' and `Barycentric Dynamical Time'
(Fairhead \& Bretagnon 1990). Finally, $\Delta_{S\odot}$ describes the
Shapiro delay in the gravitational field of the Sun (Shapiro 1964).

For binary pulsars the timing formula (\ref{sptf}) has to be extended by terms
representing orbital motion and light propagation effects in the binary system
(Blandford \& Teukolsky 1976, Damour \& Deruelle 1986, Damour \& Taylor 1992):
\be\label{bptf}\ba{l}
   T=t_{obs}+\Delta_C-D/f^2+\Delta_{R\odot}+\Delta_{E\odot}+\Delta_{S\odot}
   \\[2mm]
   \hspace*{35mm} +\Delta_R+\Delta_E+\Delta_S \,,
\ea\ee
where the major effect is the Roemer delay $\Delta_R$ which depends on the
orbital motion of the pulsar and the orientation of the pulsar orbit with
respect to the line of sight. If the binary motion is purely Keplerian then
the Roemer delay depends on 5 (Keplerian) parameters:
\begin{itemize}
\item $P_b$, the orbital period of the binary system,
\item $x\equiv a_p\sin i/c$, the projected semi-major axis,
\item $e$, the eccentricity of the orbit,
\item $\omega$, the longitude of periastron,
\item $T_0$, the time of periastron passage.
\end{itemize}
$a_p$ is the semi-major axis of the pulsar orbit, $i$ is the inclination of
the orbital plane with respect to the line of sight, where $i=90\degr$ implies
edge on, and $c$ is the speed of light. The Roemer delay caused by the
Keplerian motion of a binary system is given by
\be\label{Roem1}
   \Delta_R=x[(\cos U-e)\sin\omega+(1-e^2)^{1/2}\sin U\cos\omega] \,,
\ee 
where $U$, the eccentric anomaly, is related to time, $T$, by the Kepler
equation
\be\label{Keq}
   U-e\sin U = 2\pi\;\frac{T-T_0}{P_b} \,.
\ee

Soon after the discovery of PSR B1913+16 (Hulse \& Taylor 1975) it was clear
that a pure Keplerian timing model is not appropriate to analyse the timing
observation of this 7.8-hour orbital-period binary pulsar. For this purpose
Blandford \& Teukolsky (1976) derived a timing model (BT model) which
contains the largest short-period relativistic effect, the `Einstein delay'
$\Delta_E$, a combination of special-relativistic time dilation and
gravitational redshift.  They also included secular drifts of the main orbital
parameters by following linear-in-time expressions:
\be
   P_b = P_{b0} + \dot P_b (T-T_0) \,,
\ee\be
   x   = x_0 + \dot x (T-T_0) \,, \label{xdotBT} 
\ee\be
   e   = e_0 + \dot e (T-T_0) \,,
\ee\be
   \omega = \omega_0 + \dot\omega (T-T_0) \label{omdotBT} \,.
\ee
Based on a remarkably simple analytic solution of the post-Newtonian two-body
problem (Damour \& Deruelle 1985) Damour \& Deruelle (1986) derived an
improved timing formula (DD model) for relativistic binary pulsars. The DD
model differs from the BT model in two ways: it contains the Shapiro delay
$\Delta_S$, which is of particular importance for $i$ close to $90\degr$, and
it allows for periodic effects in the orbital motion, e.g.\ in the BT model
only the secular drift of the longitude of periastron is taken into account
(equation (\ref{omdotBT})), whereas the DD model describes both the secular
and quasi-periodic changes in $\omega$ according to
\be\label{omdotDD}
    \omega = \omega_0+kA_e(U) \,, \qquad 
    \mbox{with}\quad\dot\omega=2\pi k/P_b \,,
\ee 
where
\be\label{Ae}
   A_e(U)=2\arctan\left[\left(\frac{1+e}{1-e}\right)^{1/2}
            \tan\frac{U}{2}\right] \,,\\
\ee
and $U$ is a function of $T$ given by the solution of the (generalised)
Kepler equation
\be
   U-e\sin U = 2\pi\left[\left(\frac{T-T_0}{P_b}\right)
         -\frac{\dot P_b}{2}\left(\frac{T-T_0}{P_b}\right)^2\right] \,.
\ee
$\dot P_b$ accounts for any secular change in the orbital period, like the
damping caused by tidal dissipation or the emission of gravitational waves.  
The post-Keplerian Roemer delay is given by
\be\label{Roem2}
\ba{l}
   \Delta_R=x\{[\cos U-e(1+\delta_r)]\sin\omega(U) \\[2mm]
   \qquad\qquad +[1-e^2(1+\delta_f)^2]^{1/2}\sin U\cos\omega(U)\} \,,
\ea\ee
where $\omega(U)$ is given by equation (\ref{omdotDD}). The post-Keplerian
parameters $\delta_r$ and $\delta_f$ represent periodic post-Newtonian changes 
in the orbital motion, i.e.\ periodic changes of order $(v/c)^2$ where 
$v$ is a typical orbital velocity of the binary star system.

Damour and Taylor (1992) defined an improved version of the BT model (BT+
model) where equation (\ref{omdotBT}) is replaced by equation (\ref{omdotDD}).
The advantage of the BT+ model is that it contains the same number of
parameters as the BT model but is able to account for quasi-periodic changes
in $\omega$.

To date, more than 50 radio pulsars are known which are members of a binary
star system. The vast majority of these binary pulsars have a compact
degenerate companion which is either a helium white dwarf or a neutron star.
Timing observations for most of these binary pulsars can be fully explained by
the timing models above. In a first approximation these models can be used
for timing observations of the two main-sequence star binary pulsars, PSRs
B1259--63 and J0045--7319, by fitting for the five Keplerian parameters, and
for $\dot\omega$ and $\dot x$. However, this approximation models only the
short-term secular changes of the binary orbit correctly.

In this paper an improved timing model for binary radio pulsars with
main-sequence star companions is presented. The new timing formula accounts
for the short-term secular precessional effects, for most of the short-term
periodic orbital effects and for the long-term secular effects which are
caused by classical spin-orbit coupling. In Section 2 a detailed investigation
of the orbital dynamics of binary systems with classical spin-orbit coupling
is given. First, a simple analytic solution is presented for the case that the
orbital motion takes place in the equatorial plane of the massive
companion. This solution will be helpful when developing the new timing
formula.  Then, the orbital dynamics of the general case, i.e.\ arbitrary
orientation of the orbit with respect to the companion, is studied.  In
Section 3 it is shown how the orbital dynamics influences the timing
observation of the main-sequence star binary pulsars PSRs B1259--63 and
J0045--7319. As a consequence the new timing formula is developed.  In Section
4 the long-term validity of the old and new timing formulae is
studied. Extensions which are quadratic in time and take into account
long-term precessional effects in the binary orbit are presented. It is shown
that observations of such long-term secular effects have the potential to
probe the internal structure of the companion. In Section 5 I reinvestigate
the orbital precession of PSR J0045--7319 since results presented so far in
the literature were based on an incorrect formula for the precession of the
longitude of periastron.  In Section 6 the conclusions are given.


\section{The orbital motion of main-sequence star binary pulsars}

The quadrupole of a main-sequence star companion is given by the difference
between the moments of inertia about the spin-axis, $I_3^*$, and an orthogonal
axis, $I_1^*$. It is proportional to the mass of the companion, $m_*$, the 
(polar) radius of the rotating star, $R_*$, squared, and to the spin squared
(Cowling 1938, Schwarzschild 1958):
\be
   I_3^*-I_1^* = \mfrac{2}{3}k\,m_*R_*^2\hat\Omega_*^2 \equiv m_*q \,,
\ee
where $k$ is the apsidal motion constant and $\hat\Omega_* \equiv \Omega_*/
(Gm_*/R_*^3)^{1/2}$ is the dimensionless proper rotation of the companion.
$\Omega_*$ is the angular velocity of the companion's proper rotation. A
main-sequence star of $10M_\odot$ has $k\la0.03$, $R_*\sim6R_\odot$. If the
star is rotating at 70\% of its break-up velocity, which appears to be a
typical value for Be stars (Porter 1996), one finds for its quadrupole moment
\be\label{q}
   q = \mfrac{2}{3}kR_*^2\hat\Omega_*^2 \la 0.35\:R_\odot^2
     = 7.6\times10^{-6}\:{\rm AU}^2 \,.
\ee

The spin-induced quadrupole moment of the companion leads to an additional
$1/r^3$ potential term in the gravitational interaction between the two
components which implies an apsidal motion and, when the spin of the companion
is not aligned with the orbital angular momentum, to a precession of the
orbital plane. The general expressions for the rates of apsidal motion and
orbital precession can be found in Smarr \& Blandford (1976) and Kopal (1978),
(see Section 4 in this paper). The expressions in Smarr \& Blandford (1976)
and Kopal (1978) are derived by averaging the orbital dynamics over a full
orbital period, in order to get the secular changes in the orbit. This
procedure, by definition, neglects all short-term periodic effects. But, as
will be shown in this paper, for main-sequence star binary pulsars with a long
orbital period and a high eccentricity, like PSRs B1259--63 and J0045--7319,
these short-term periodic effects are important.

For a study of the short-term periodic effects one needs the orbital motion in
full detail. In the centre-of-mass system the (Newtonian) orbital dynamics of
a binary pulsar with an oblate companion star is given by the Hamiltonian
(Barker \& O'Connell 1975)
\be\label{H}
    \mcal{H} = \frac{\mbf{p}^2}{2\mu}-\frac{GM\mu}{r}\left(1
    +\frac{q}{2r^2}\left[1-3(\hat\mbf{s}\cdot\hat\mbf{n})^2\right]\right) \,,
\ee
where the linear momentum $\mbf{p}$ is related to the linear momenta of pulsar
and companion by $\mbf{p}=\mbf{p}_p=-\mbf{p}_*$. $\mbf{r}$ is a vector
pointing from the companion to the pulsar, $r\equiv|\mbf{r}|$,
$\hat\mbf{n}\equiv\mbf{r}/r$, $M=m_p+m_*$ is the total mass of the system,
$\mu\equiv m_pm_*/M$ is the reduced mass and $\hat\mbf{s}$ is the unit vector
in direction of the spin of the companion.

Before studying the full dynamics, I will investigate the special case of
motion in the equatorial plane. Although the two known main-sequence star
binary radio pulsars do not orbit their companion in the equatorial plane the
comparably simple solution of this problem will be helpful in developing a new
timing formula in the next section.

\subsection{The equatorial motion}

For motion in the equatorial plane ($\hat\mbf{s}\cdot\hat\mbf{n}=0$) the
Hamiltonian (\ref{H}) reduces to
\be
    \mcal{H}_{\perp} = \frac{\mbf{p}^2}{2\mu}
           -\frac{GM\mu}{r}\left(1+\frac{q}{2r^2}\right) \,.
\ee
The invariance of this Hamiltonian under time translation and spatial
rotations implies the conservation of the (reduced) energy, $E\equiv
\mcal{H}_{\perp}/ \mu$, and the (reduced) total angular momentum, $\mbf{J}
\equiv\mbf{r}\times \mbf{p}/\mu$. If one introduces polar coordinates, $\mbf{r}
=r(\cos f,\sin f,0)$, and makes use of the conserved quantities one finds the
following equations of motion:
\be
   \dot r^2 = 2E+\frac{2GM}{r}-\frac{J^2}{r^2}+\frac{GMq}{r^3} \,, 
\ee\be
   \dot f = \frac{J}{r^2} \,,
\ee
where $J\equiv|\mbf{J}|$. These equations of motion can be solved to first
order in $q$ by a quasi-Keplerian trigonometric parametrisation (cf.\ Damour
\& Deruelle 1985). For bound orbits, $E<0$, one finds
\be
    2\pi\frac{t-t_0}{P_b} = U - e_t\sin U \,,
\ee\be
    r = a(1-e_r\cos U) \,,
\ee\be
    f-f_0 = (1+k)A_{e_f}(U) \,,
\ee
where $A_{e_f}(U)$ is given by equation (\ref{Ae}) and
\be
   P_b = \frac{2\pi GM}{(-2E)^{3/2}} \,,
\ee\be
   e_t = \left(1+\frac{2EJ^2}{G^2M^2}-\frac{2Eq}{J^2}\right)^{1/2} \,,
\ee\be
   a = -\frac{GM}{2E}\left(1+\frac{Eq}{J^2}\right) \,,
\ee\be
   e_r \equiv (1+\delta_r)e_t = \left(1-\frac{Eq}{J^2}\right)e_t \,,
\ee\be
   k = \frac{3}{2}\left(\frac{GM}{J^2}\right)^2q \,,
\ee\be
   e_f \equiv (1+\delta_f)e_t = \left(1-\frac{2Eq}{J^2}\right)e_t \,.
\ee
The motion of the pulsar is
\be
   \mbf{r}_p = \frac{m_*}{M}\:r\:(\cos f,\sin f,0) \,. 
\ee
Therefore, if the orbital angular momentum and the spin of the companion are
(nearly) aligned then the orbital motion of the binary system can be described
to first order in $q$ by a simple trigonometric parametrisation which is
identical to the `quasi-Keplerian' parametrisation given by Damour \& Deruelle
(1985) to solve the post-Newtonian two-body problem. Further, the timing
observations of such a system can be fully explained using the DD timing
model.

\subsection{The general case}

The full dynamics given by the Hamiltonian (\ref{H}) is a well known problem
in the theory of Earth satellite motion. Due to the anisotropic nature of the
quadrupole potential one does not expect any simple analytic solution as in
the previous section.  Various methods have been developed to solve this
problem (see e.g.\ Hagihara 1970, Roy 1978).  To first order in $q$ the
dynamics given by equation (\ref{H}) can be solved by the {\it method of the
variation of the elements} (osculating orbits). The following six elements
are used to represent the osculating orbit:
\begin{itemize}
\item $a$, the semi-major axis of the (relative) orbit
\item $e$, the eccentricity of the orbit
\item $\theta$, the inclination of the orbital plane
\item $\phi$, the longitude of the ascending node 
\item $\psi$, the longitude of periastron
\item $\mcal{M}$, the mean anomaly
\end{itemize}
The angles are defined with respect to the equatorial plane of the oblate
companion (see Fig.~1).

\begin{figure}
\psfig{figure=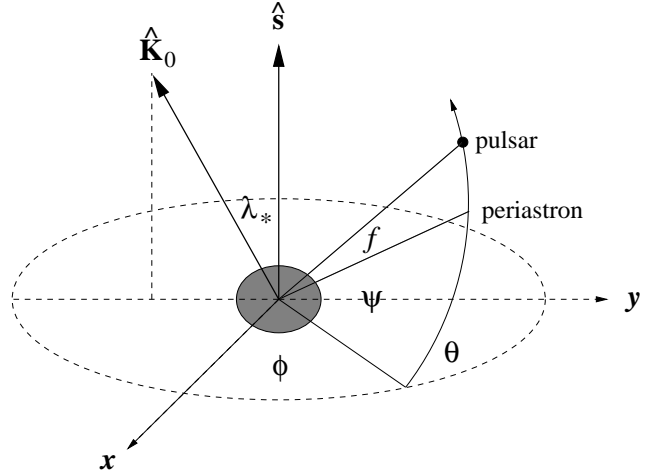,angle=-90,width=84mm}
\caption{Definition of the angles $\theta$, $\phi$, and $\psi$ with respect to
the equatorial coordinate system. $f$ is the true anomaly and $\lambda_*$ is
the angle between the line-of-sight and the spin of the companion.}
\end{figure}

\noindent
Lagrange's planetary equations for this problem can be written in the form
\be\label{ve1}
   \frac{da}{dt}=\frac{2}{na}\:\frac{\partial\mcal{R}}{\partial\mcal{M}} \,,
\ee\be
   \frac{de}{dt}=\frac{(1-e^2)^{1/2}}{na^2e}\left(
   (1-e^2)^{1/2}\frac{\partial\mcal{R}}{\partial\mcal{M}}
   -\frac{\partial\mcal{R}}{\partial\psi}\right) \,,
\ee\be
   \frac{d\theta}{dt}=\frac{1}{na^2(1-e^2)^{1/2}\sin\theta}\left(\cos\theta\:
   \frac{\partial\mcal{R}}{\partial\psi}-\frac{\partial\mcal{R}}
   {\partial\phi}\right) \,,
\ee\be
   \frac{d\phi}{dt}=\frac{1}{na^2(1-e^2)^{1/2}\sin\theta}\:
   \frac{\partial\mcal{R}}{\partial\theta} \,,
\ee\be
   \frac{d\psi}{dt}=\frac{1}{na^2(1-e^2)^{1/2}}\left(
    \frac{1-e^2}{e}\frac{\partial\mcal{R}}{\partial e}
    -\cot\theta\:\frac{\partial\mcal{R}}{\partial\theta}\right) \,,
\ee\be\label{ve6}
   \frac{d\mcal{M}}{dt}=n-\frac{2}{na^2}\left(\frac{1-e^2}{2e}\:
   \frac{\partial\mcal{R}}{\partial e}+a\frac{\partial\mcal{R}}{\partial a}
   \right) \,.
\ee
The relevant parts of the disturbing function $\mcal{R}\equiv\mcal{R}_s+
\mcal{R}_p$ are
\be
   \mcal{R}_s = \frac{q n^2}{2}\left(1-\frac{3}{2}\sin^2\theta\right)
      (1-e^2)^{-3/2} \,, 
\ee\be\ba{l}
   \DS\mcal{R}_p = \frac{q n^2}{2}\left(\frac{a}{r}\right)^3\left[\left
     (1-\frac{3}{2}\sin^2\theta\right)\left(1-\frac{(a/r)^3}
     {(1-e^2)^{3/2}}\right)\right.\\[3mm] \DS
     \hspace*{39mm}\left.+\frac{1}{2}\sin^2\theta\cos 2(\psi+f)
     \frac{}{}\right] \,.
\ea\ee
$\mcal{R}_s$ and $\mcal{R}_p$ are first-order secular and short-period parts
respectively of the disturbing function $\mcal{R}$.

The secular perturbations of the first order are obtained by putting
$\mcal{R}=\mcal{R}_s$ in equations (\ref{ve1}) to (\ref{ve6}). The result is
\be
   \Delta a_s=0 \,, \quad \Delta e_s=0 \,, \quad \Delta\theta_s=0 \,,
\ee\be\label{phisec}
   \Delta\phi_s = -Q\:\cos\theta_0\:n_0t \,,
\ee\be\label{psisec}
   \Delta\psi_s = Q\left(2-\frac{5}{2}\sin^2\theta_0\right)n_0t \,,
\ee\be
   \Delta\mcal{M}_s = \left[1+Q\left(1
      -\frac{3}{2}\sin^2\theta_0\right)(1-e_0^2)^{1/2}\right]\:n_0t\,,
\ee
where the definition
\be
   Q \equiv \frac{3q}{2p_0^2} \quad (\ll 1)
\ee
was used. The zero subscript indicates evaluation at the initial
condition. $n_0$ is related to the initial semi-major axis, $a_0$, by
\be
   n_0^2 a_0^3 = GM \,.
\ee

To simplify the representation of the first-order short-periodic perturbations
I define the functions
\be\ba{l}
   \DS S_{e,\theta;f} \equiv\left(1-\frac{3}{2}\sin^2\theta\right) \\[2mm]
   \DS \qquad\quad\times\left[\left(1-\frac{e^2}{4}\right)\sin f
       +\frac{e}{2}\sin 2f+\frac{e^2}{12}\sin 3f\right] \,,\\[4mm]
   \DS C_{e,\theta;f} \equiv\left(1-\frac{3}{2}\sin^2\theta\right) \\[2mm]
   \DS \qquad\quad\times\left[\left(1+\frac{e^2}{4}\right)\cos f
       +\frac{e}{2}\cos 2f+\frac{e^2}{12}\cos 3f\right] \,,
\ea\ee
and
\be\ba{l}
   S_{\psi;f}(\alpha_0,\alpha_1,\alpha_2,\alpha_3,\alpha_4,\alpha_5)\equiv \\
   \qquad\qquad\qquad\DS 
   \alpha_0\sin(2\psi-f)+\sum_{m=1}^5\alpha_i\sin(2\psi+mf) \,,\\
   C_{\psi;f}(\alpha_0,\alpha_1,\alpha_2,\alpha_3,\alpha_4,\alpha_5)\equiv \\
   \qquad\qquad\qquad\DS 
   \alpha_0\cos(2\psi-f)+\sum_{m=1}^5\alpha_i\sin(2\psi+mf) \,,
\ea\ee
and
\be
   W_{e;f} \equiv f-\mcal{M}+e\sin f \,.
\ee
To derive the first-order short-period perturbations, the disturbing function
$\mcal{R}$ in equations (\ref{ve1}) to (\ref{ve6}) is replaced by
$\mcal{R}_p$.  Integration of the resulting equations leads to the following
expressions for the six elements (Kozai 1959, Fitzpatrick 1970):
\be\ba{l}\label{d1}
   \Delta a_p = \DS Q\frac{2e_0a_0}{1-e_0^2}\left\{C_{e_0,\theta_0;f}
   \right.\\[2mm]\left.\quad
   +\sin^2\theta_0\:C_{\psi_0,f}\left(\mfrac{e_0^2}{16},\mfrac{12+3e_0^2}{16},
   \mfrac{2+3e_0^2}{4e_0},\mfrac{12+3e_0^2}{16},\mfrac{3e_0}{8},
   \mfrac{e_0^2}{16}\right)\right\} \,,
\ea\hspace*{-10mm}\ee
\be\ba{l}
   \Delta e_p = \DS Q\left\{C_{e_0,\theta_0;f} 
   \right.\\[2mm]\left.\quad
   +\sin^2\theta_0\:C_{\psi_0;f}\left(\mfrac{e_0^2}{16},\mfrac{4+11e_0^2}{16},
   \mfrac{5e_0}{4},\mfrac{28+17e_0^2}{48},\mfrac{3e_0}{8},
   \mfrac{e_0^2}{16}\right)\right\} \,,
\ea\hspace*{-10mm}\ee
\be\ba{l}
   \Delta\theta_p = Q\cos\theta_0\sin\theta_0\: 
   C_{\omega_0;f}\left(0,\frac{e_0}2,\frac{1}2,\frac{e_0}6,0,0\right) \,,
\ea\ee
\be\ba{l}
   \Delta\phi_p = Q\cos\theta_0\left\{-W_f+S_{\psi_0;f}
   \left(0,\frac{e_0}2,\frac{1}2,\frac{e_0}6,0,0\right)\right\} \,,
\ea\ee
\be\ba{l}
   \Delta\psi_p = \DS \frac{Q}{e_0}\left\{\left(2-\frac{5}{2}
   \sin^2\theta_0\right)e_0 W_f+S_{e_0,\theta_0;f}
   \right.\\[2mm]\left.\quad
   -e_0\:S_{\psi_0;f}\left(0,\frac{e_0}2,\frac{1}2,\frac{e_0}6,0,0\right)
   \right.\\[2mm]\left.\quad
   +\sin^2\theta\:S_{\psi_0;f}\left(-\mfrac{e_0^2}{16},
   -\mfrac{4-15e_0^2}{16},\mfrac{5e_0}{4},\mfrac{28+19e_0^2}{48},
   \mfrac{3e_0}{8},\mfrac{e_0^2}{16}\right)\right\} \,,
\ea\hspace*{-10mm}\ee
\be\ba{l}\label{d6}
   \Delta\mcal{M}_p = \DS -Q\frac{(1-e_0)^{1/2}}{e_0}\left\{S_{e_0,\theta_0;f}
   \right.\\[2mm]\left.\quad +\sin^2\theta_0\:
   S_{\psi_0,f}\left(-\mfrac{e_0^2}{16},-\mfrac{4+5e_0^2}{16},0,
   \mfrac{196-7e_0^2}{336},\mfrac{3e_0}{8},\mfrac{e_0^2}{16}\right)\right\}\,.
\ea\hspace*{-10mm}\ee

The mean and true anomaly are connected by the equations
\be\label{anom1}
   \mcal{M} = U - e\sin U \,,
\ee\be\label{anom2}
    f = 2\arctan\left[\left(\frac{1+e}{1-e}\right)^{1/2}
                \tan\frac{U}{2}\right] \,.
\ee
The distance between the main-sequence star and the pulsar is 
\be
    r = a (1-e\cos U) = \frac{a(1-e^2)}{1+e\cos f} \,.\\
\ee
The position vector of the pulsar at the time $t$ with respect to the
equatorial coordinate system is
\bea\label{rvec}
   \mbf{r} = r\left(
   \ba{rrr}
   \cos\phi & -\sin\phi & 0 \\
   \sin\phi &  \cos\phi & 0 \\
   0 & 0 & 1 
   \ea \right)\left(
   \ba{l}
   \cos(\psi+f) \\
   \sin(\psi+f)\cos\theta \\
   \sin(\psi+f)\sin\theta
   \ea
   \right) \,.
\eea
where the (variable) parameters $a,e,\theta,\phi,\psi,\mcal{M}$  are given by
\be
   \xi(t) = \xi_0 + [\Delta\xi_s(t)-\Delta\xi_s(0)]
                  + [\Delta\xi_p(t)-\Delta\xi_p(0)] 
\ee
($\xi$ stands for one of these six parameters). To calculate $\Delta\xi_p$ one
uses $\mcal{M}=\mcal{M}_0+n_0t$, $e=e_0$ in equations (\ref{anom1}) and
(\ref{anom2}) to obtain an approximated value for $f$ which then is used in
evaluating equations (\ref{d1}) to (\ref{d6}).

As an example I substitute the orbital elements of the main-sequence star
binary pulsar PSR B1259--63 (Table~1) into equations (\ref{d1}) to (\ref{d6}),
For the initial value of $\theta$ ($\theta_0$) I take $60\degr$, which is a
realistic value for this binary system (see Wex et al.\ 1997).  The values for
$\phi_0$ and $\psi_0$ can be estimated from $\omega$, $i$, and $\lambda_*$:
\be\label{A8}
   \cos\phi =  
         \frac{\cos i-\cos\theta\cos\lambda_*}{\sin\theta\sin\lambda_*}\,,
\ee\be
   \sin(\omega-\psi) = -\frac{\sin\lambda_*\sin\phi}{\sin i}\,,
\ee\be
   \cos(\omega-\psi) = 
   \frac{\sin\theta\cos\lambda_*-\cos\theta\sin\lambda_*\cos\phi}{\sin i}\,.
\ee
(Equation (\ref{A8}) does not determine $\phi$ uniquely, since $\phi$ runs
between 0 and $2\pi$.) $\lambda_*$ can be derived from optical observations of
the projected proper rotation of the Be star. Johnston et al.\ (1994) found
$\lambda_*$ $\sim$ $30\degr$ (or $150\degr$).  For the `strength' of the
quadrupole I use $q=2\times10^{-6}\:{\rm AU}^2$ ($k\sim0.01$, cf.\ equation
(\ref{q})).

The results for two full orbits of PSR B1259--63 are given in Fig.~2 and
Fig.~3. Fig.~2 presents the changes of the four `periodic' parameters,
$a,e,\theta,\mcal{M}$, which take their initial value after every full
period. Fig.~3 presents the changes of $\phi$ and $\psi$. I call $\phi$ and
$\psi$ `secular' parameters since they show both periodic and secular
changes. It is obvious that for all parameters the major changes take place
very close to the periastron passages, as expected from the $1/r^3$ nature of
the quadrupole potential.

\begin{figure}
\vspace*{-0.4cm}
\psfig{figure=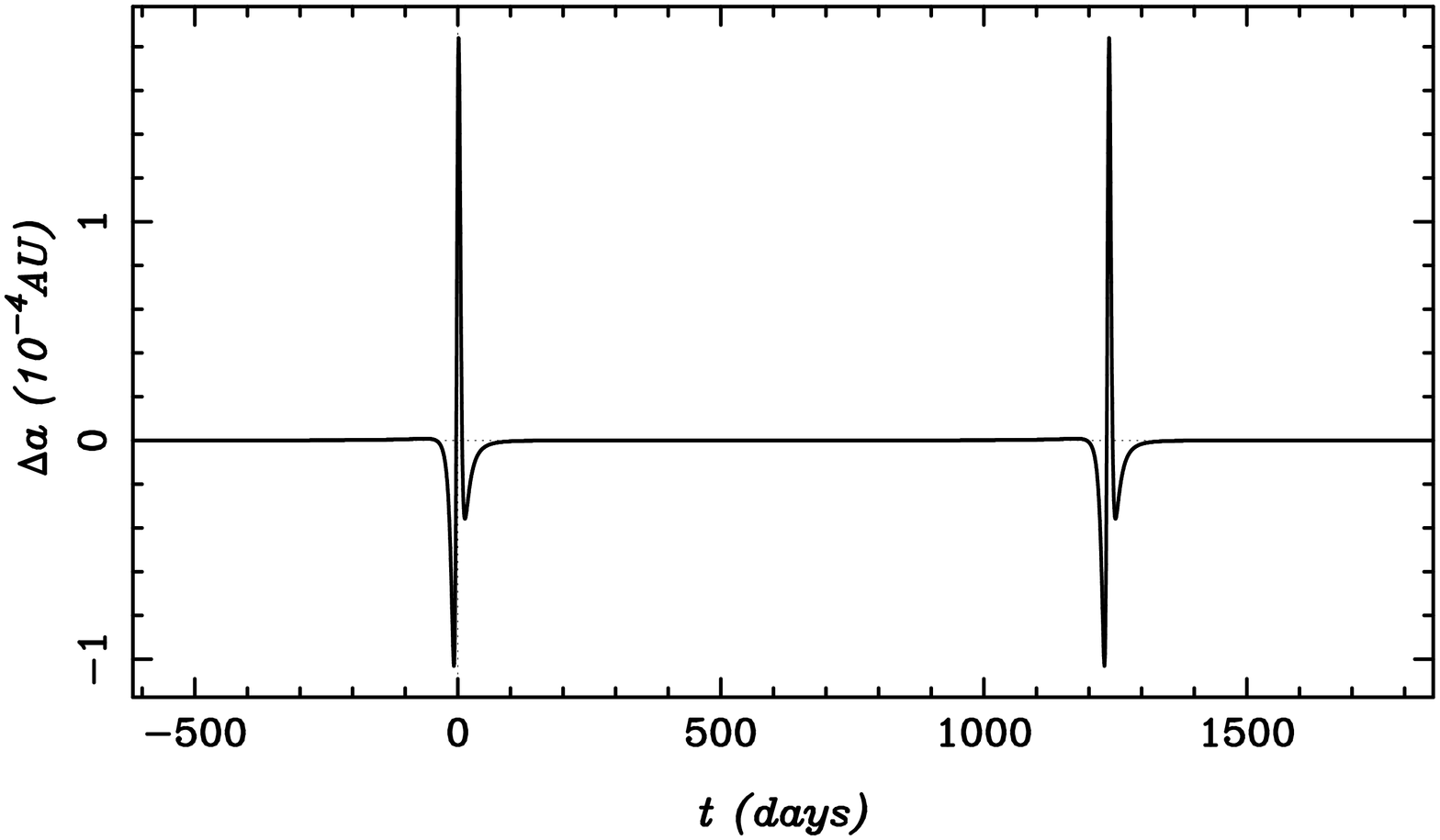,width=8.5cm,angle=-90}
\vspace{-0.8cm}
\psfig{figure=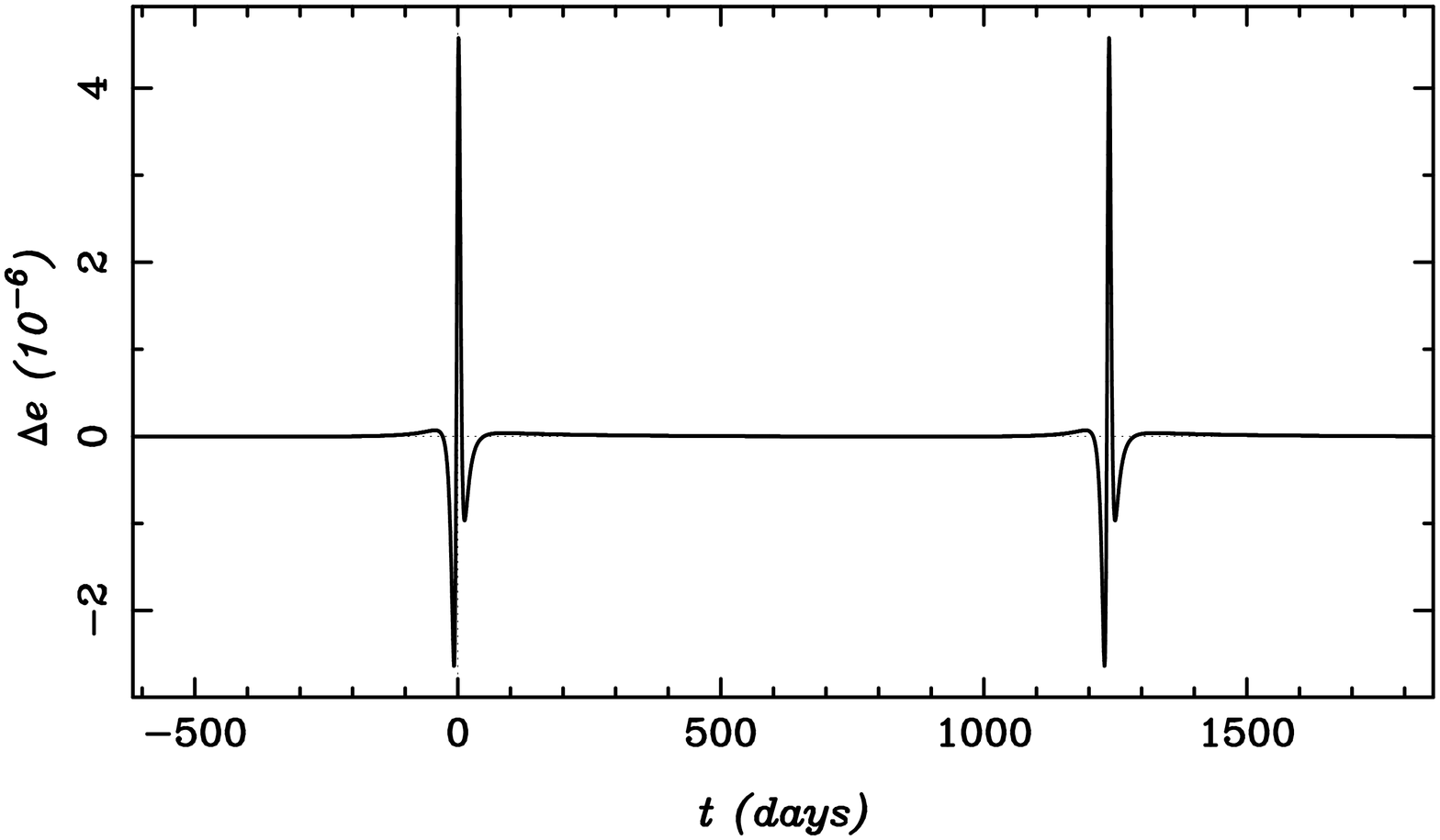,width=8.5cm,angle=-90}
\vspace{-0.8cm}
\psfig{figure=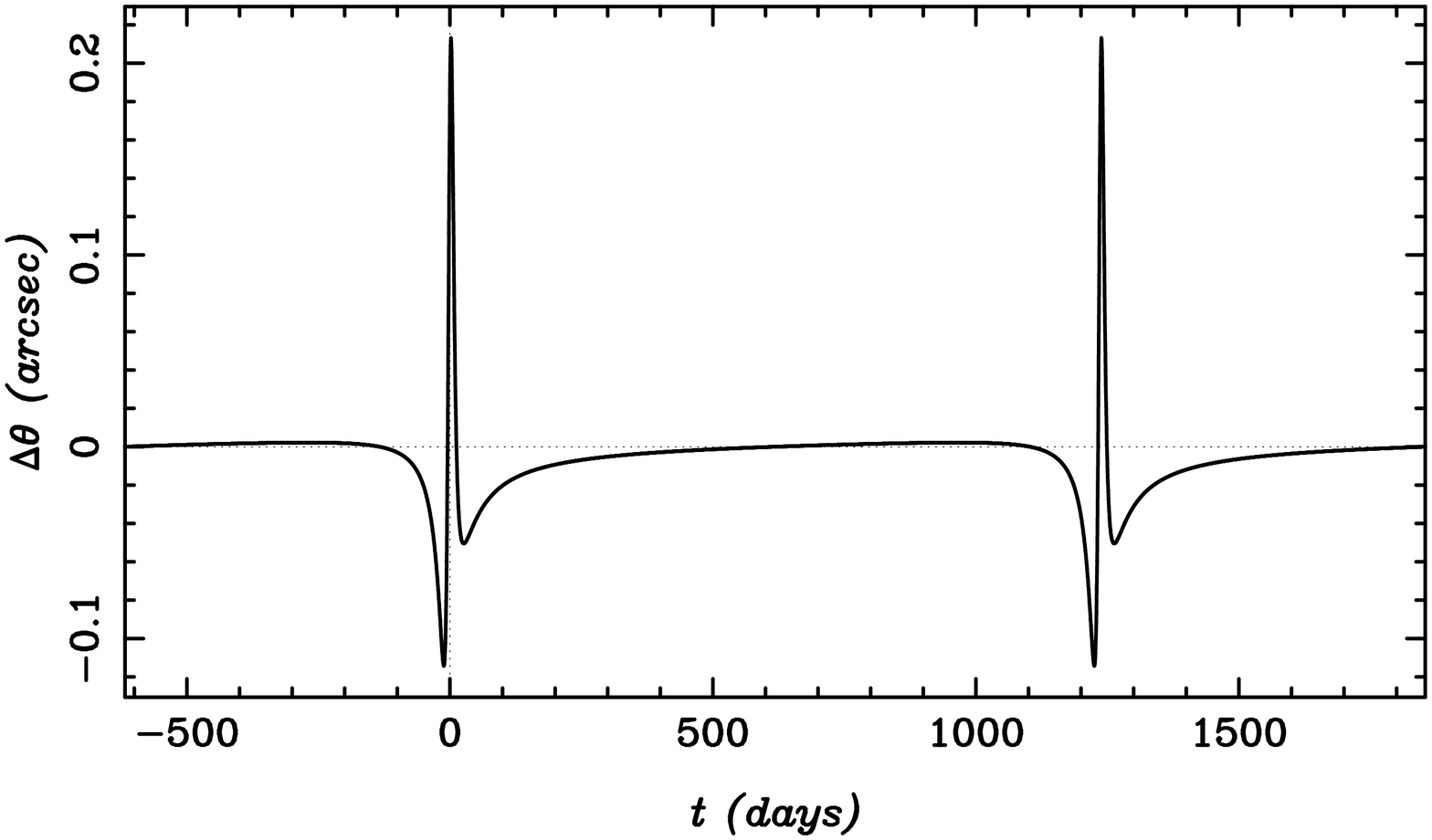,width=8.5cm,angle=-90}
\vspace{-0.8cm}
\psfig{figure=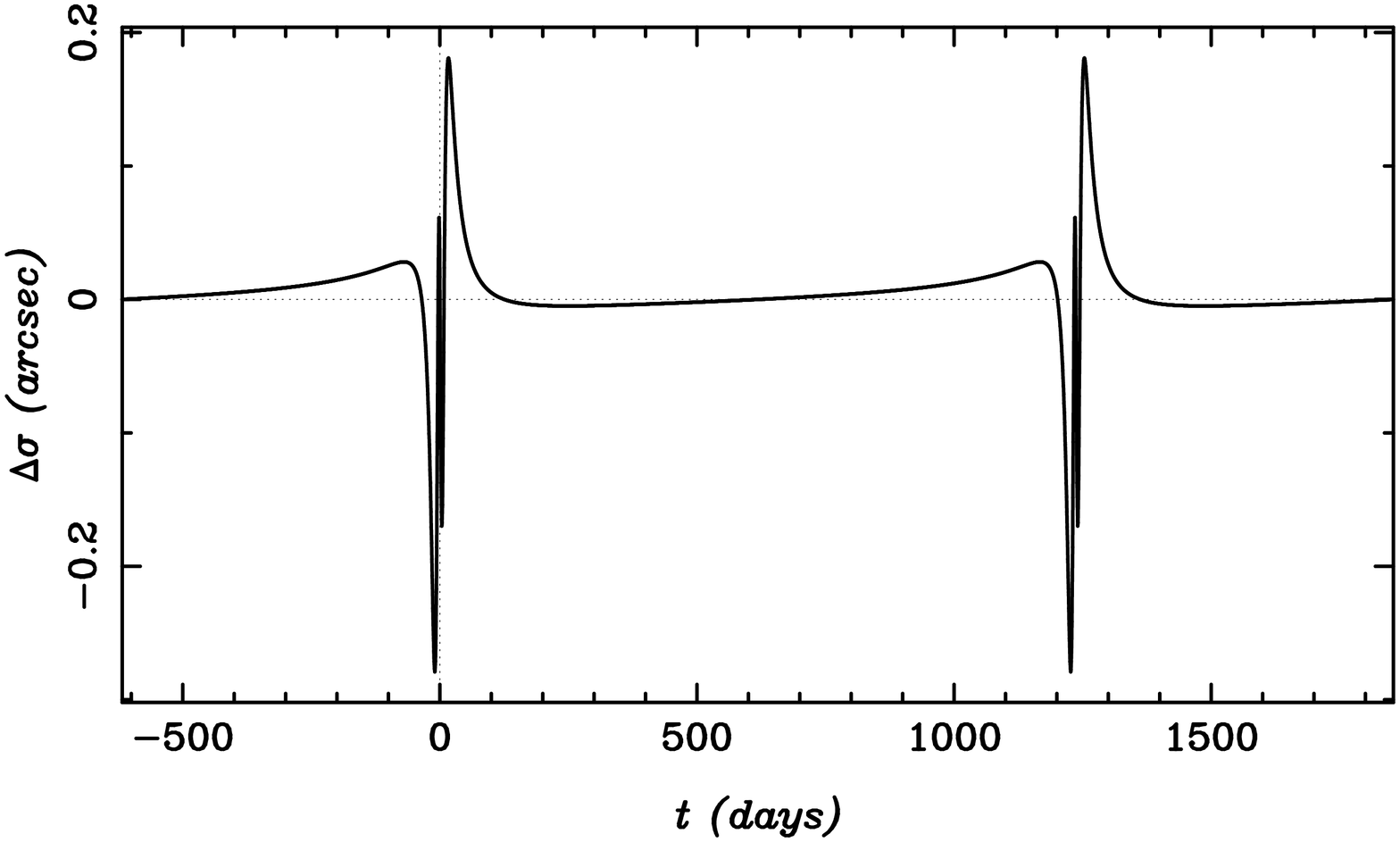,width=8.5cm,angle=-90}
\vspace{-0.8cm}
\caption{Calculated changes of the `periodic' parameters of PSR B1259--63.
I used the parameters of Table~1 ($i_0=36\degr$), $\lambda_*=30\degr$,
$\theta_0=60\degr$, $\phi_0=30\degr$, $\psi_0=114\degr$, and
$q=2\times10^{-6}\:{\rm AU}^2$.}
\end{figure}

\begin{figure}
\vspace*{-0.4cm}
\psfig{figure=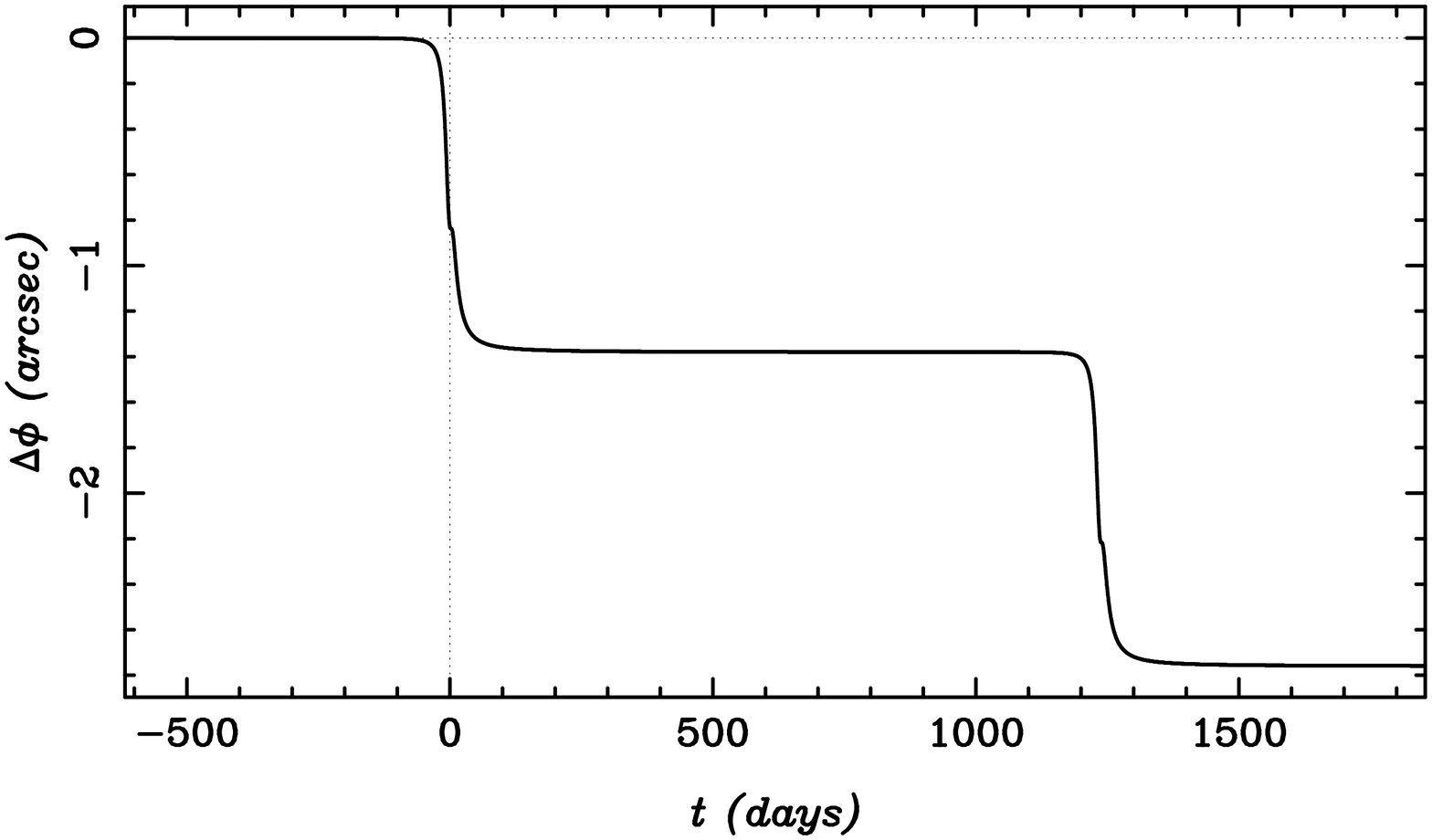,width=8.5cm,angle=-90}
\vspace{-0.8cm}
\psfig{figure=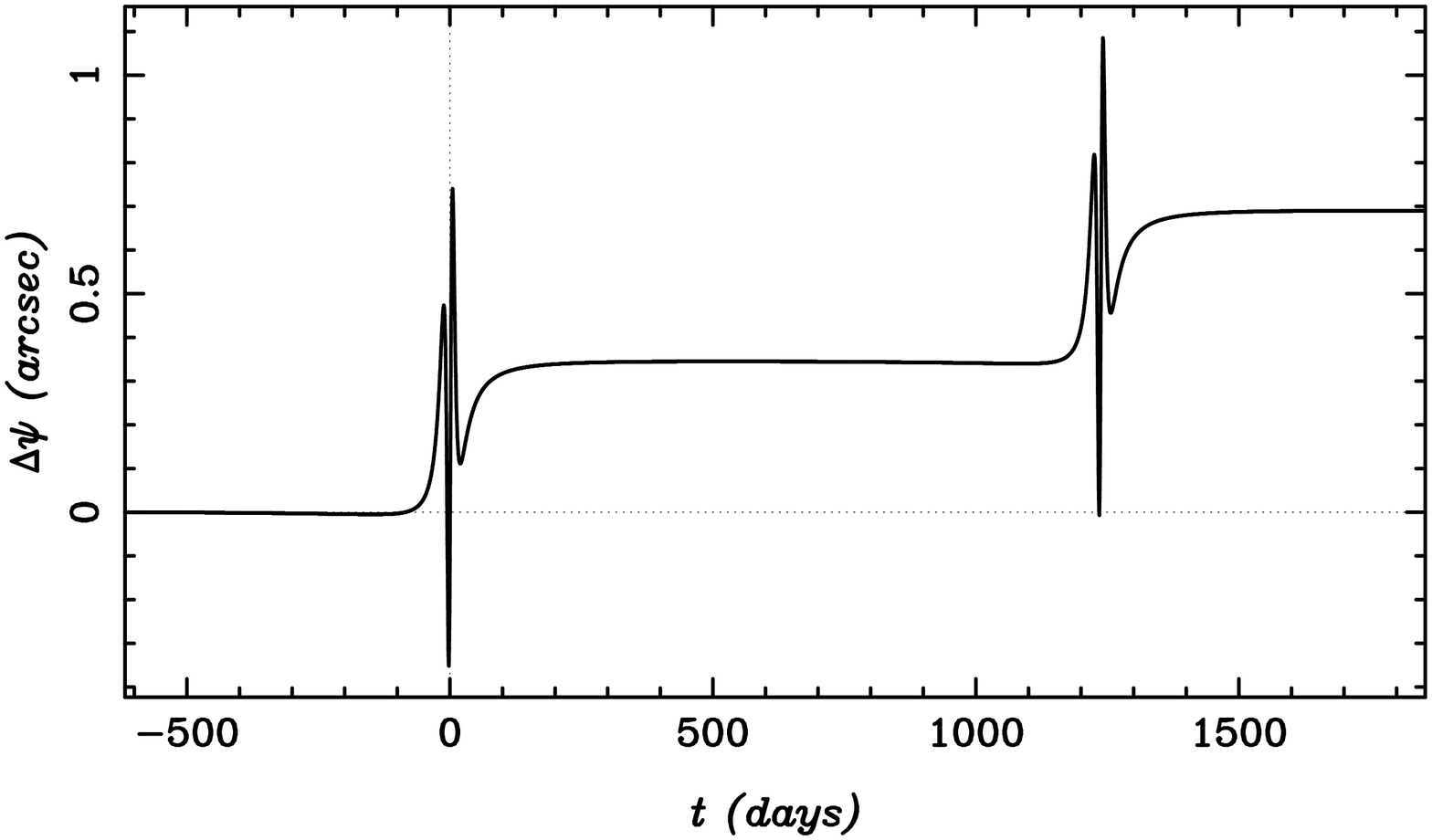,width=8.5cm,angle=-90}
\vspace{-0.8cm}
\caption{Calculated changes of the `secular' parameters of PSR B1259--63.
(Parameters as in Fig.~2.)}
\end{figure}


\section{Timing models for main-sequence star binary pulsars I. Short-term
periodic effects}

Let $\hat\mbf{K}_0=(0,-\sin\lambda_*,\cos\lambda_*)$ be the unit vector which
indicates the direction of sight (see Fig.~1). The Roemer delay measured by an
observer on Earth is then given by
\be
   \Delta_R = \frac{1}{c}\:\hat\mbf{K}_0\cdot\mbf{r}_p \,,
\ee
where $c$ is the speed of light and 
\be
   \mbf{r}_p = \frac{m_*}{M}\mbf{r}
\ee
is the position vector of the pulsar originating in the centre of mass of the
binary system. Using equation (\ref{rvec}) leads to
\be\ba{l}
   \Delta_R=\DS\frac{m_*}{M}\:\frac{r}{c}\:\{-\sin\lambda_*
            [\sin\phi\cos(\psi+f) \\[2mm] 
   \quad +\cos\phi\sin(\psi+f)\cos\theta]+\cos\lambda_*\sin(\psi+f)
   \sin\theta\} \,.
\ea\ee

As mentioned in the introduction, the simplest timing model for binary pulsars
is the BT model where changes in the longitude of periastron, $\omega$, and
changes in the projected semi-major axis of the pulsar orbit, $x$, are assumed
to be linear in time. The discussion in the previous section clearly showed
that the (osculating) parameters of a binary system with an oblate companion do
not change linearly in time (cf.\ Fig.~2 and Fig.~3). Thus one does not expect
that the application of the BT model leads to a perfect fit. Fig.~4 shows the
difference between equations (\ref{Roem1}),(\ref{xdotBT}),(\ref{omdotBT}) and
the actual Roemer delay expected for the binary pulsars PSRs B1259--63 and
J0045--7319. (Using the DD model instead leads to similar results). The
typical precision in the measurement of the arrival time of pulsar signals is
of the order of 100 $\mu$s for PSR B1259--63 (Johnston et al.\ 1994) and a few
ms for PSR J0045--7319 (Kaspi et al.\ 1994). For both pulsars the deviations
given in Fig.~4 are larger than the error in the TOAs. For PSR B1259--63 it is
more than a factor of ten. Therefore the BT model is only a very crude
approximation for these two binary pulsars and there is the need for an
improved timing model, for PSR B1259--63 in particular.

\begin{figure}
\vspace*{-0.5cm}
\psfig{figure=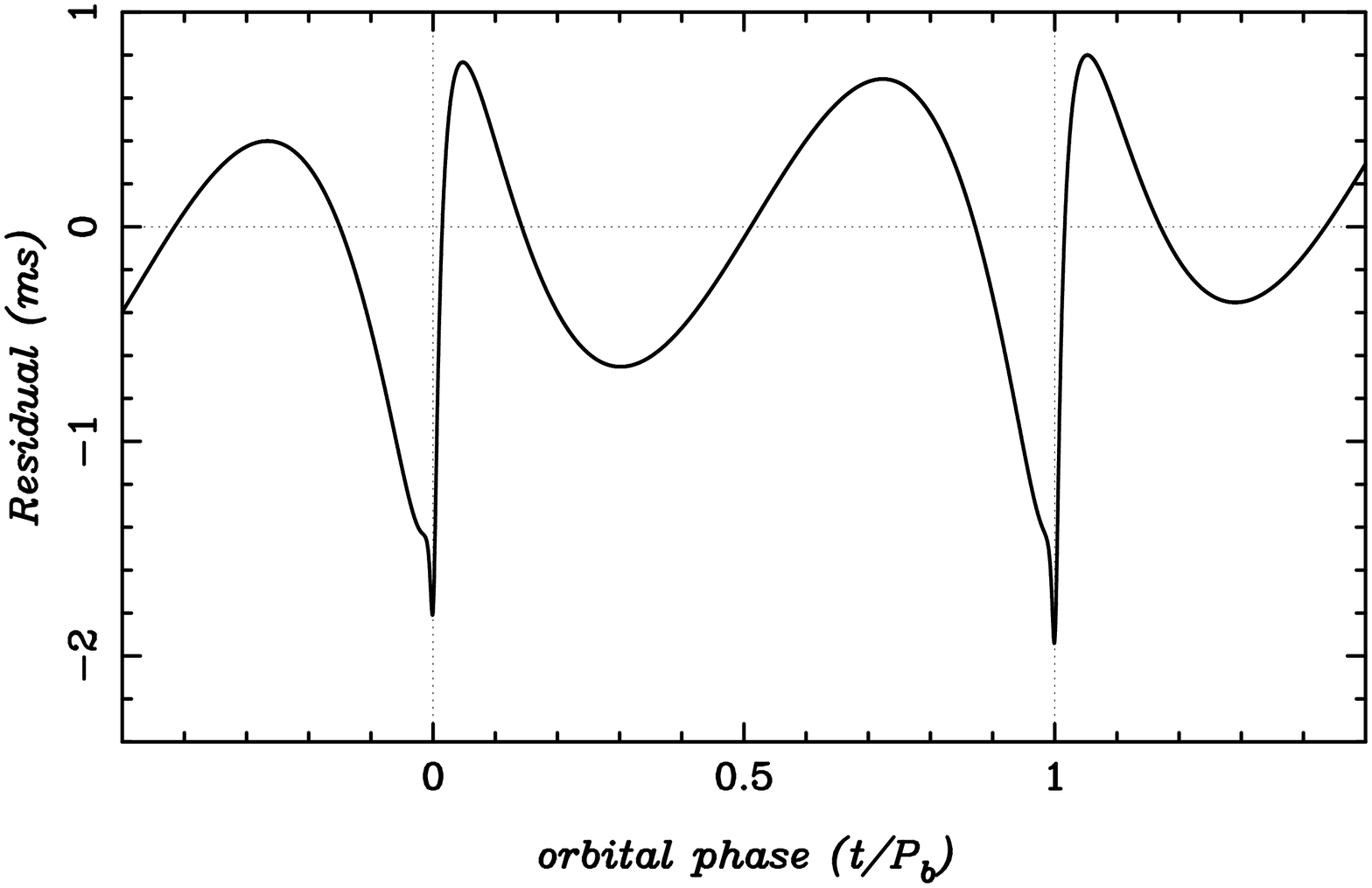,width=8.5cm,angle=-90}
\vspace{-0.5cm}
\psfig{figure=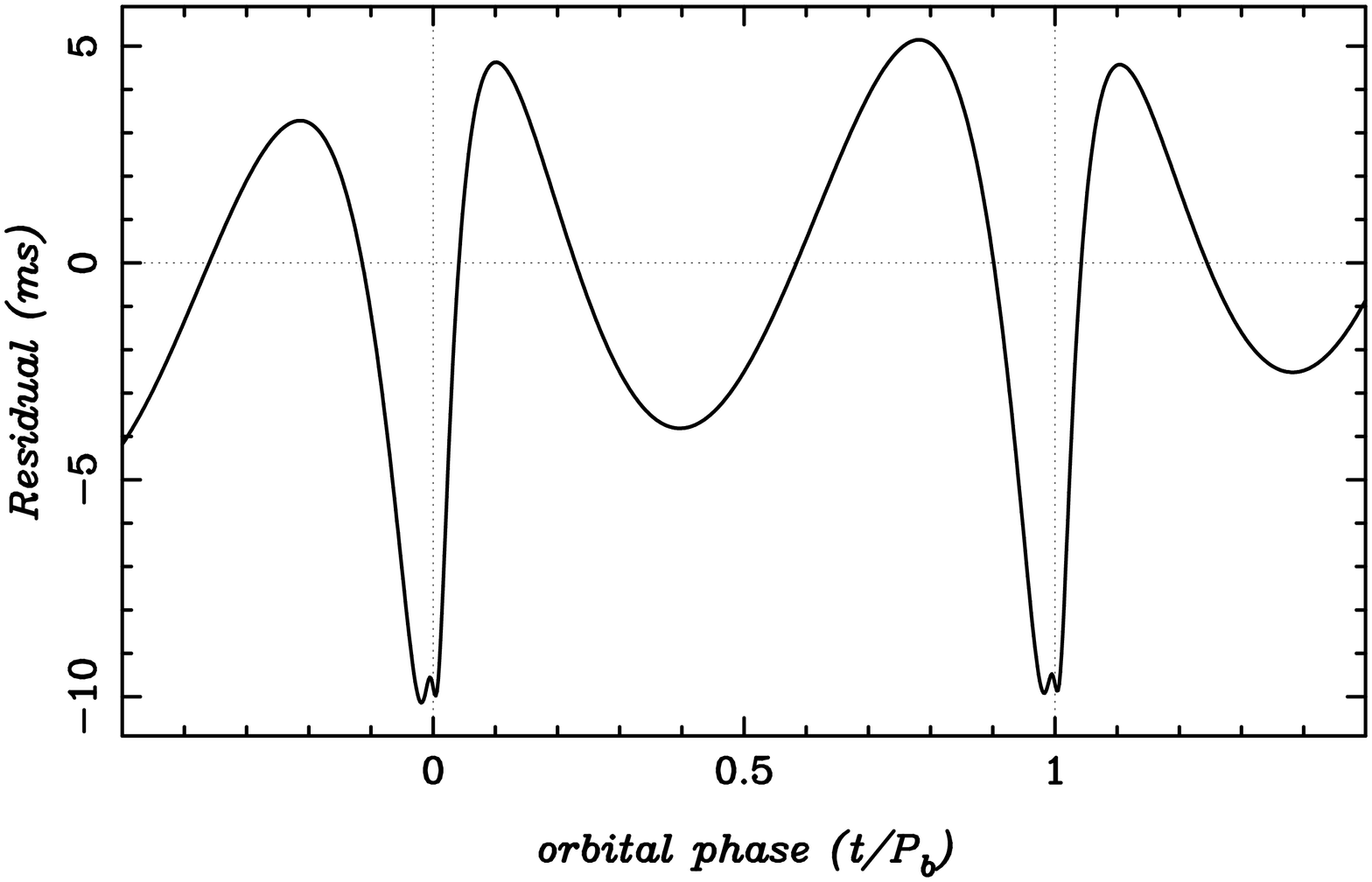,width=8.5cm,angle=-90}

\caption{Roemer delay as used in the BT timing model minus Roemer delay as
expected for PSRs B1259--63 (upper figure) and J0045--7319 (lower figure).
For B1259--63 I used the same parameters as in Fig.~2. For J0045--7319 I used
the parameters of Table~1 ($i_0=44\degr$), $\lambda_*=15\degr$,
$\theta_0=30\degr$, $\phi_0=156\degr$, $\psi_0=106\degr$, and
$q=2\times10^{-6}\:{\rm AU}^2$.}
\end{figure}

To extract reliable information from timing observations of main-sequence star
binary pulsars one could construct a timing model that contains the full
orbital dynamics given in the previous section. In the (unlikely) case that
the orbital motion takes place in the equatorial plane of the Be star one can
use the DD timing model as shown by the solution in section 2.1. For the
general case one has to use the solution of Section 2.2 which leads to a
rather complicated timing formula with a comparably high number of parameters
to fit for where some of these parameters are only indirectly related with
observable effects. Therefore, given the limited number of TOAs and the finite
size of their measurement errors, one sees that in general a timing formula
based on the equations of Section 2.2 is not a practical procedure. What one
is looking for is a simple timing model which is a very good approximation to
reality. In the ideal case the number of parameters should be the same as in
the BT model.

The main deficit of the BT model is the use of equations (\ref{xdotBT}) and
(\ref{omdotBT}) to describe the precession of the orbit. Even for a purely
equatorial motion (see Section 2.1) equation (\ref{omdotBT}) has to be
replaced by (\ref{omdotDD}) according to the DD timing model. If the orbit is
tilted with respect to the equatorial plane the precession of the orbit leads
to a change in the inclination of the orbital plane with respect to the line
of sight. Fig.~5 gives this change of $i$ for PSR B1259--63 as a function of
the time, $t$, and as a function of the true anomaly, $f$. The change of $i$
is neither linear in $t$ nor linear in $f$. But the assumption of linearity in
$f$ is obviously much closer to reality than the assumption of linearity in
$t$. The same is true for the change of $x$ which is a function of $i$.

\begin{figure}
\psfig{figure=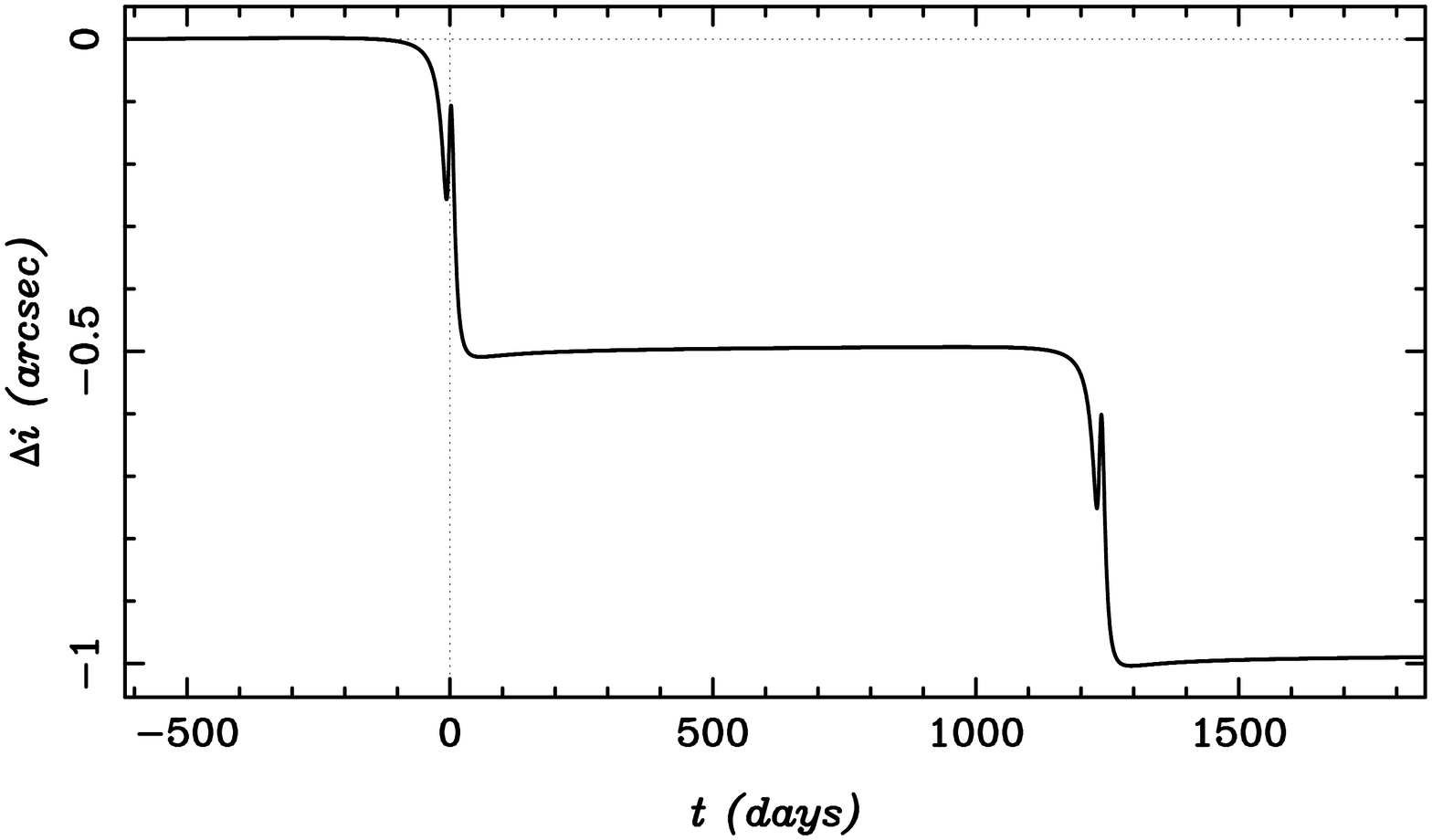,width=8.5cm,angle=-90}
\vspace{-0.8cm}
\psfig{figure=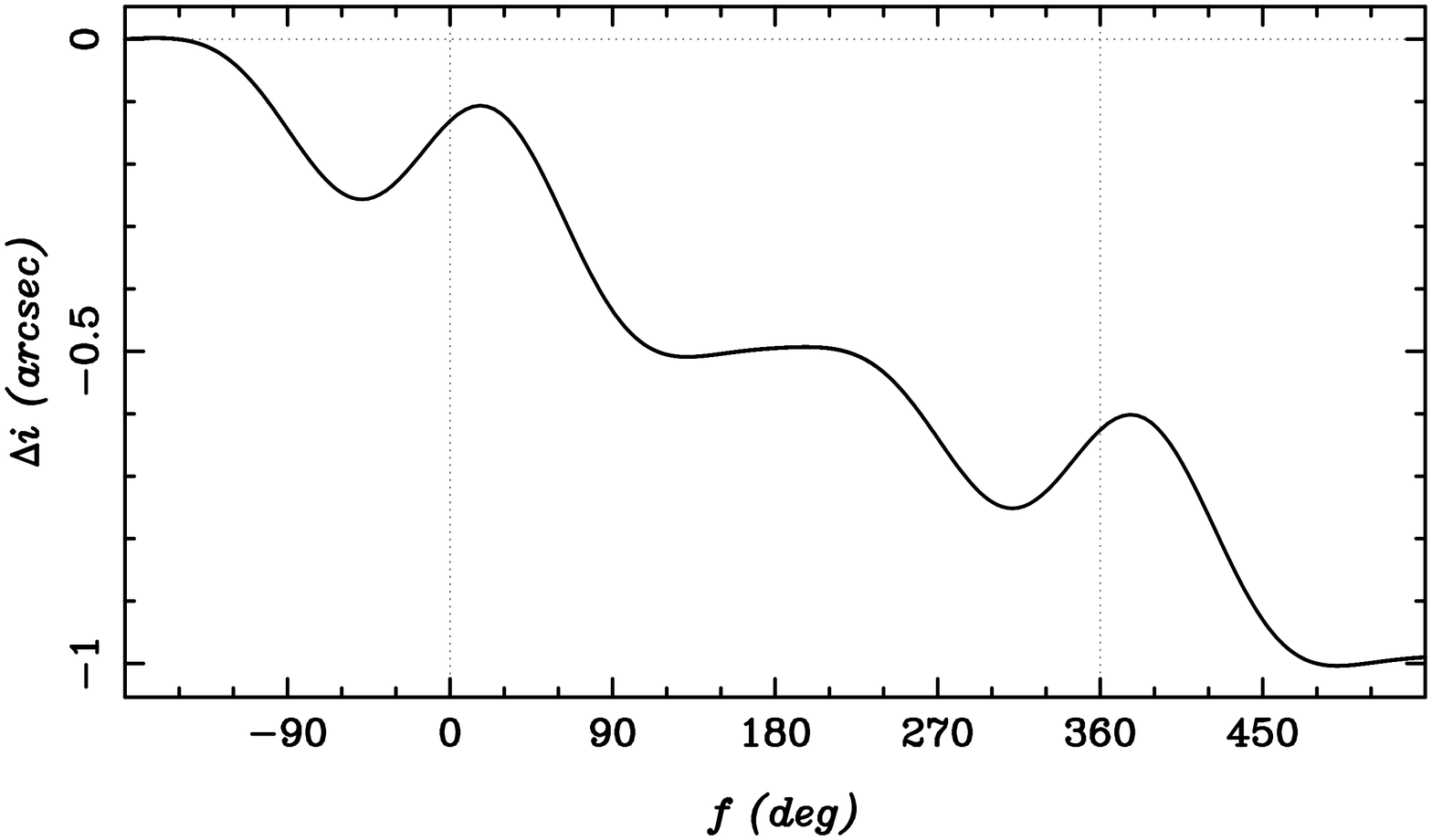,width=8.5cm,angle=-90}
\vspace{-0.8cm}
\caption{Calculated change of the inclination $i$ of the orbit of PSR
B1259--63 as function of the time, $t$, (upper figure) and of the true
anomaly, $f$, (lower figure). (Parameters as in Fig.~2)}
\end{figure}

\noindent
Therefore I define a new timing model which I call BT++ model.  Analogue to
the construction of the BT+ model by Damour and Taylor (1992) the BT++ model
is defined by replacing equations (\ref{xdotBT}) and (\ref{omdotBT}) in the BT
model by
\be\label{xdot}
    x = x_0 + \xi A_e(U)\,,
    \qquad \dot x \equiv 2\pi\xi/P_b\,,
\ee
and
\be\label{omdot}
    \omega = \omega_0 + k A_e(U)\,,
    \qquad \dot\omega \equiv 2\pi k/P_b\,.
\ee  
A comparison of the expected Roemer delay and the Roemer delay as used in the
BT++ model is given in Fig.~6. For both main-sequence star binary pulsars,
PSRs B1259--63 and J0045--7319, the BT++ model is off by clearly less than
the typical error in the TOAs. Only very close to periastron the deviations
for PSR B1259--63 show a sharp peak of 200 $\mu$s, a value which is slightly
larger then the typical measurement error. On the other hand, so far there are
no timing observation of this pulsar close to periastron. The reason is the
occultation by the circumstellar disk which lasts from $\sim$20 days before
until $\sim$20 days after periastron (Johnston et al.\ 1996).  

I conclude that one should certainly use the BT++ model instead of the BT
model to fit the TOAs of PSR B1259--63.  The BT++ model has the same number of
parameters as the BT model, but is able to account for the fact that
changes of the binary parameters happen mainly close to periastron. The
BT++ model was already applied successfully to fit the TOAs of PSR B1259--63
(Wex et al.\ 1997). I expect a slight improvement of the residuals for PSR
J0045--7319 when one uses the BT++ model.

\begin{figure}
\vspace*{-0.5cm}
\psfig{figure=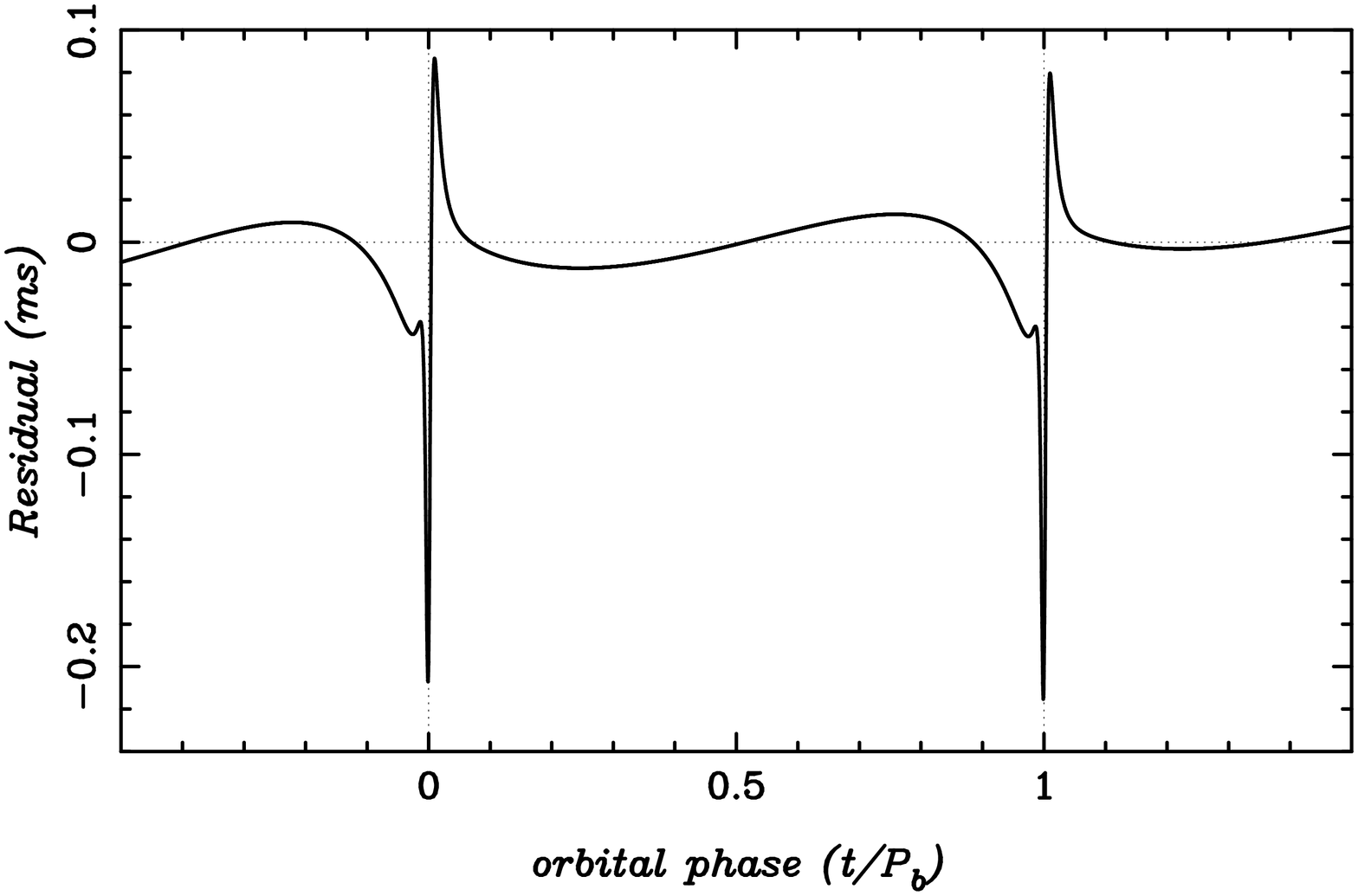,width=8.5cm,angle=-90}
\vspace{-0.5cm}
\psfig{figure=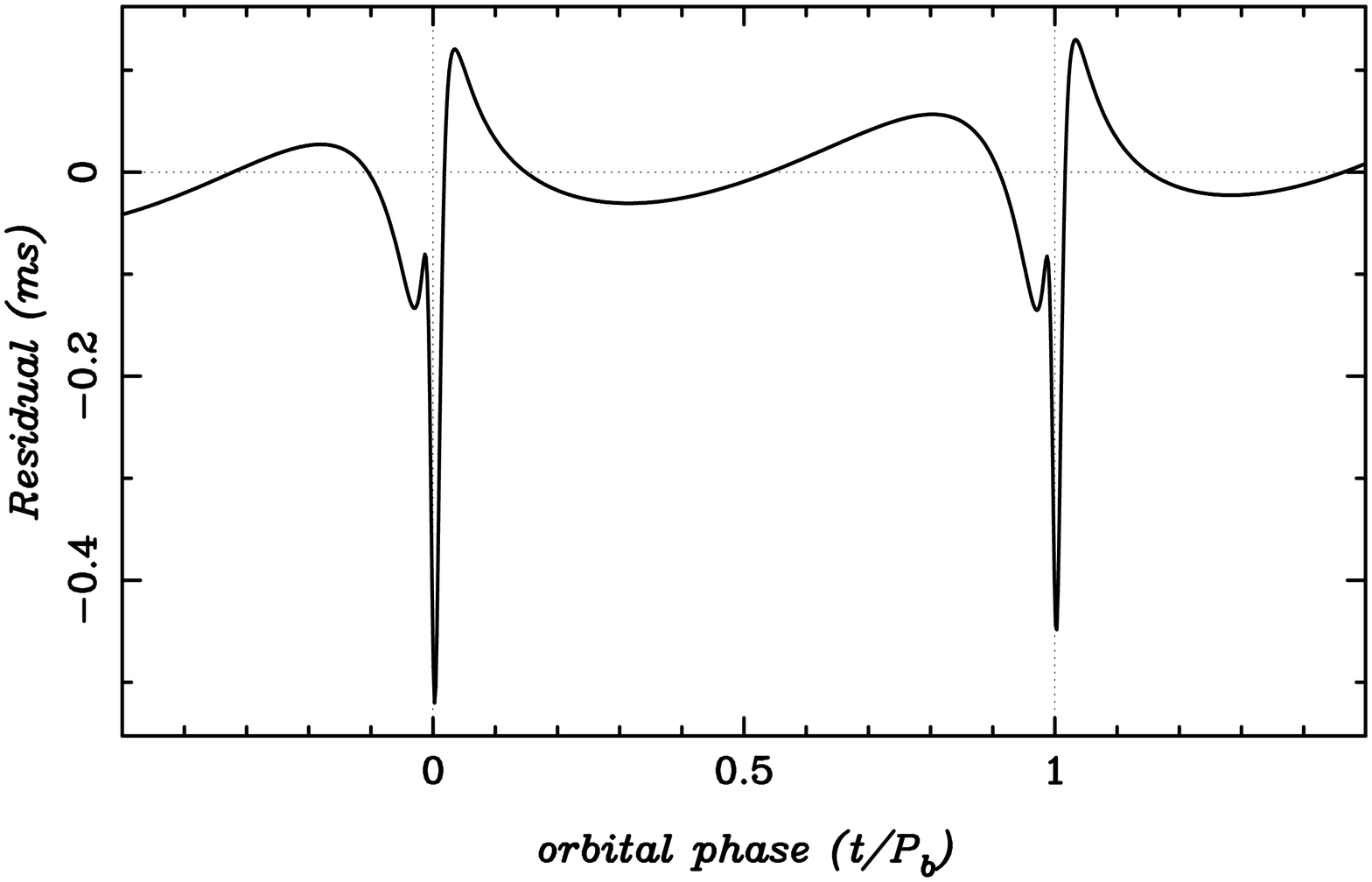,width=8.5cm,angle=-90}
\caption{Roemer delay as used in the BT++ timing model minus Roemer delay as
expected for PSRs B1259--63 (upper figure) and J0045--7319 (lower figure).
(Parameters as in Fig.~4.)}
\end{figure}

As concluded in Section 2.1, the correct timing model for equatorial orbits is
the DD model. The DD model takes into account all the periodic effects of the
(equatorial) orbital motion. One can now try to construct an even better
timing model for main-sequence star binary pulsars, say DD+, by replacing
equation (\ref{xdotBT}) in the DD model by equation (\ref{xdot}).  The result
is a timing model which combines the advantages of the BT++ model in
describing the precession of the orbital plane and the DD model in describing
periodic orbital effects. The representation of the Roemer delay in the DD+
model contains one more parameter than in the BT++ model. The DD+ model has
the same number of parameters as the DD model. From Fig.~7 one sees that the
DD+ model represents a nearly perfect fit for most parts of the orbit and
close to periastron it is an improvement by a factor of 2 with respect to the
BT++ model. At present the measurement precision for the TOAs for PSRs
B1259--63 and J0045--7319 does not allow to fit for the (full) DD+ model.

\begin{figure}
\vspace*{-0.5cm}
\psfig{figure=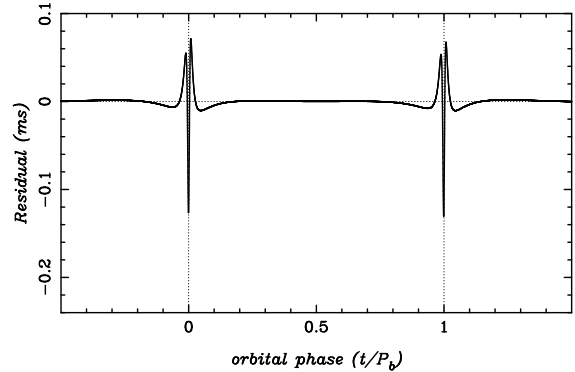,width=8.5cm,angle=-90}
\caption{Roemer delay as used in the DD+ timing model minus Roemer delay as
         expected for PSRs B1259--63. (Parameters as in Fig.~2.)}
\end{figure}

Finally, the upper figure of Fig.~5 indicates that a step-function model
($\omega$ and $x$ change discontinuously at each periastron) will also give a
good approximation to reality. In fact, Fig.~8 shows that a step-function
model is clearly better than the BT model (lower figure of Fig.~4). On the
other hand, comparison between the lower figure of Fig.~6 and Fig.~8 implies
that a step-function model is clearly worse than the BT++ model, in particular
for observations close to periastron.

\begin{figure}
\vspace*{-0.5cm}
\psfig{figure=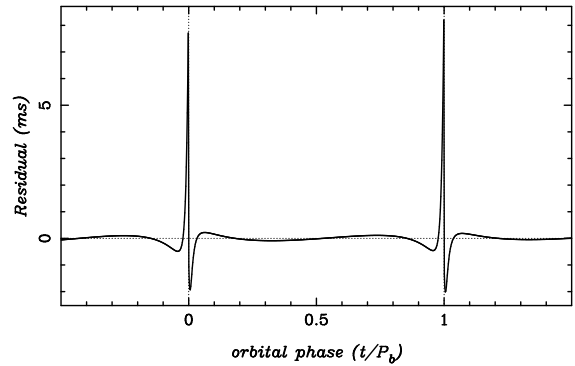,width=8.5cm,angle=-90}
\caption{Roemer delay as used in the step-function timing model minus Roemer
         delay as expected for PSRs J0045--7319. (Parameters as in Fig.~4.)}
\end{figure}


\section{Timing models for main-sequence star binary pulsars II. Long-term
secular effects}

So far only the short-term effects in the orbital motion have been dealt
with. I have shown the advantage of the BT++ (and DD+) model in taking into
account the short-term periodic effects of the orbital motion.  In the BT,
BT+, DD, BT++ and DD+ model the secular changes in $\omega$ and $x$ are
assumed to be linear in time.  This approximation will hold as long as there
are only small changes in $\omega$ and $x$.  In this section I will focus on
the long term precession of the binary orbit and its influence on pulsar
timing and will investigate the limits of the present timing models.

In the following discussion I neglect periodic effects and focus only on the
secular changes in the orbit of the binary system caused by the spin induced
quadrupole of the main-sequence star companion. The solution presented in
Section 2.2 does not give the long term behaviour correctly. It does not take
into account the change in the orientation of the main-sequence star due to
spin-orbit coupling. The change of the orientation of the main-sequence star
is of order $q$ and thus appears in the equations of motion at order $q^2$
which was neglected in Section 2.2. On long time scales the orientation of the
main-sequence star changes by a comparably large amount and therefore the
contribution, although of order $q^2$, becomes numerically significant. The
solution in Section 2.2 is perfectly suited for a discussion of periodic
effects during a few orbital turns, but for the study of the long-term
behaviour one should focus on the conserved quantities, which are the total
energy and the total angular momentum. The total angular momentum, $\mbf{J}$,
is the sum of the orbital angular momentum, $\mbf{L}$, and the spin of the
main-sequence star, $\mbf{S}$. On average, over one full period, the length of
$\mbf{S}$, $|\mbf{S}|$, and $\mbf{L}$, $|\mbf{L}|$, are conserved (Barker \&
O'Connel 1975).  Fig.~9 illustrates the resulting orbital dynamics.

\begin{figure}
\psfig{figure=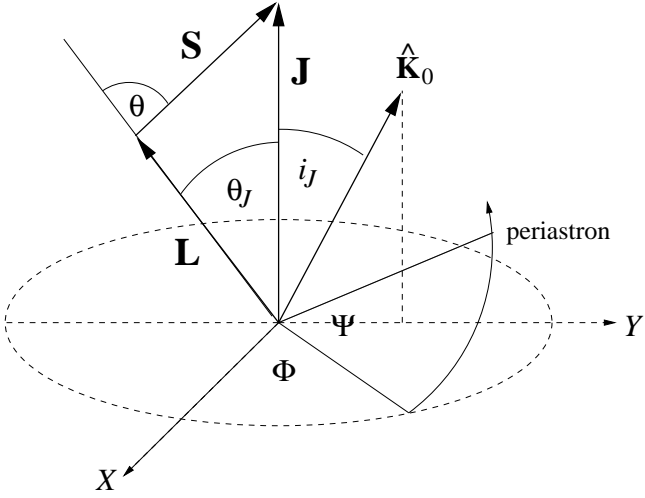,width=8.5cm,angle=-90}
\caption{Definition of various angles in the total-angular-momentum reference
	frame. The invariable ($X$-$Y$) plane is perpendicular to the total
	angular momentum $\mbf{J}=\mbf{L}+\mbf{S}$. The line-of-sight
	($\mbf{K}_0$) is in the $Y$-$Z$ plane. $\mbf{J}$ is a conserved
	quantity and, if averaged over one full period, $|\mbf{L}|$ and
	$|\mbf{S}|$ are conserved quantities. Thus $i_J$, $\theta_J$, and
	$\theta$ are conserved. The angles $\Phi$ and $\Psi$ change linearly
	with time.}
\end{figure}

Averaged over one orbital period one finds for the change of $\Phi$ and 
$\Psi$ (Smarr \& Blandford 1976, Kopal 1978)
\be\label{prec1}
   \bar{\dot\Phi} = -\bar n\bar Q \left(\frac{\sin\bar\theta\cos\bar\theta}
   {\sin\bar\theta_J}\right)=const. 
\ee
and
\be\label{prec2}
   \bar{\dot\Psi} = \bar n\bar Q \left(1-\frac{3}{2}\sin^2\bar \theta+
   \frac{1}{2}\sin 2\bar\theta\cot\bar\theta_J\right)=const.\,,
\ee
where
\be
   \bar Q = \frac{3(I^*_3-I^*_1)/m_*}{2\bar a^2(1-\bar e^2)^2}
          = \frac{kR_*^2\hat\Omega^2_*}{\bar a^2(1-\bar e^2)^2} \,.
\ee
The bar on top of the quantities indicates that they are averaged over a full
orbital period (c.f.\ Section 2.2).  For simplicity I will skip the bar on top
of the averaged quantities for the rest of this section. Equations
(\ref{prec1}) and (\ref{prec2}) can be derived directly from equations
(\ref{phisec}) and (\ref{psisec}).

For the inclination of the orbit with respect to the line of sight, $i$, and
the longitude of periastron, $\omega$, one finds the relations
\be\ba{l}\label{isec}
   \cos i = \cos i_J \cos\theta_J - \sin i_J \sin\theta_J \cos \Phi \,,
\ea\ee
\be\ba{l}
   \DS\sin\omega=\frac{1}{\sin i}[(\sin\theta_J\cos i_J
   +\cos\theta_J\sin i_J\cos\Phi)\sin\Psi \\
   \hspace*{4.5cm} +\sin i_J\sin\Phi\cos\Psi) \,,
\ea\ee
\be\ba{l}\label{wsec}
   \DS\cos\omega=\frac{1}{\sin i}[(\sin\theta_J\cos i_J
   +\cos\theta_J\sin i_J\cos\Phi)\cos\Psi \\
   \hspace*{4.5cm} -\sin i_J\sin\Phi\sin\Psi) 
\ea\ee
(see Fig.~9 for the definition of $i_J$, $\theta_J$, $\Phi$ and $\Psi$).
The angles $i_J$ and $\theta_J$ are conserved quantities. The angles $\Phi$
and $\Psi$ evolve linearly in time, $t$,
\be\label{angvel}
   \Phi = \Phi_0+\dot\Phi(t-t_0) \quad\mbox{and}\quad
   \Psi = \Psi_0+\dot\Psi(t-t_0) \,.
\ee
Equations (\ref{isec}) to (\ref{angvel}) give the full (secular) evolution of
the projected semi-major axis, $x=a_p\sin i/c$, and the longitude of
periastron, $\omega$. This evolution is clearly non-linear in time since
changes in $\Psi$ couple in a complicated way with changes in $\Phi$ to
produce the secular changes in $i$ and $\omega$.

A first approximation of this non-linear behaviour can be given by
\footnote[2]{To include short-term periodic effects, as discussed in the
previous section, one has to replace $\dot x_0(t-t_0)$ by $\xi A_e(U)$ and 
$\dot\omega_0(t-t_0)$ by $kA_e(U)$; see equations 
(\ref{xdot}),(\ref{omdot}).}
\be
   x(t)\simeq x_0+\dot x_0(t-t_0)+\mfrac{1}{2}\ddot x_0(t-t_0)^2 \,,
\ee
and
\be
   \omega(t)\simeq\omega_0+\dot\omega_0(t-t_0)
   	+\mfrac{1}{2}\ddot\omega_0(t-t_0)^2 \,.
\ee
The quantities $\dot x_0$, $\ddot x_0$, $\dot\omega_0$, $\ddot\omega_0$, if
measured in timing observations, contain information about the orientation and
the quadrupole moment of the companion star.

In general the orbital angular momentum is much larger than the spin of the
companion star and thus $\theta_J\ll 1$. In case of the binary pulsars PSRs
B1259--63 and J0045--7319 $|\mbf{S}|/\mbf{L}\sim 0.1$ and $\theta_J\la
6\degr$. If $\theta_J\ll i,\pi-i$ one finds after an (Laurent) expansion of
equations (\ref{isec}) to (\ref{wsec}) with respect to the small angle
$\theta_J$:
\be
    i = i_J + \theta_J\cos\Phi + \mcal{O}(\theta_J^2) \,,
\ee\be
   \omega = \Psi+\Phi-\theta_J\cot i\sin\Phi + \mcal{O}(\theta_J^2) \,,
\ee
and
\be\label{xdotq}
   \dot x_0 = nQx_0\cot i\sin\theta\cos\theta\sin\Phi_0
                + \mcal{O}(\theta_J) \,,
\ee\be
   \ddot x_0 = -n^2Q^2x_0\cot i\left(\frac{\sin^2\theta\cos^2
          \theta\cos\Phi_0}{\sin\theta_J}\right)+\mcal{O}(\theta_J^0) \,,
\ee\be\label{omdotq}
   \dot\omega_0 = nQ\left(1-\frac{3}{2}\sin^2\theta
   +\cot i\sin\theta\cos\theta\cos\Phi_0\right)+\mcal{O}(\theta_J)\,,
\ee\be
   \ddot\omega_0 = n^2Q^2\cot i\left(\frac{\sin^2\theta\cos^2\theta
              \sin\Phi_0}{\sin\theta_J}\right) + \mcal{O}(\theta_J^0)\,.
\ee
While doing these expansions it is important
to keep in mind that $\dot\Phi$ and the leading term of $\dot\Psi$ are of
order $\theta_J^{-1}$ but $\dot\Phi+\dot\Psi$ is only of order $\theta_J^0$. A
fact which has been overlooked by Smarr and Blandford (1976) and Lai et al.\
(1995) leading to a wrong result for $\dot\omega_0$.

Since $Q$ is a quantity which is not well known, it is not possible to extract
the angles $\theta$ and $\Phi_0$ just from the measurement of $\dot x_0$ and
$\dot\omega_0$.  But, if the masses, and therefore $i$, are known within a
certain accuracy (which I assume for the following considerations) the
measurement of $\dot x_0$ and $\dot\omega_0$ restricts possible values of
$\theta$ and $\Phi_0$ to a small region in the $\theta$-$\Phi_0$ plane by
dividing equation (\ref{omdotq}) by equation (\ref{xdotq}), which leads to
\be\label{ratio}
   \frac{\dot\omega_0 x_0}{\dot x_0}\sin\Phi_0-\cos\Phi_0 = 
   \frac{1+3\cos2\theta}{2\cot i\sin 2\theta} \,.
\ee
The sign of $\dot x_0$ or $\dot\omega_0$ will help to exclude further regions
in the $\theta$-$\Phi_0$ plane. Since the right-hand sides of equations
(\ref{xdotq}) and (\ref{omdotq}) have to be positive or negative depending on
the sign of $\dot x_0$ and $\dot\omega_0$ respectively (c.f.\ Fig.~11).

If one is able to measure either $\ddot x$ or $\ddot\omega$, in addition to 
$\dot x$ and $\dot\omega$, one can make use of the following restrictions on
$\Phi_0$ and $\theta_J$:
\be\label{ratio2a}
   \frac{\dot x_0^2\tan i}{\ddot x_0x_0}=-\sin\Phi_0\tan\Phi_0\:\theta_J \,
\ee
or
\be\label{ratio2b}
   \frac{\dot x_0^2\tan i}{\ddot \omega_0x_0^2}=\sin\Phi_0\:\theta_J\,.
\ee
If one is able to measure $\dot x$, $\dot\omega$, $\ddot x$, and $\ddot\omega$
for a binary pulsar with a main-sequence star companion, then
\be
   \frac{\ddot \omega_0x_0}{\ddot x} = -\tan\Phi_0
\ee
will give $\Phi_0$. Equation (\ref{ratio2a}) or (\ref{ratio2b}) will then give
$\theta_J$ and equation (\ref{ratio}) will give $\theta$. Therefore all angles
in the geometry of the binary system are determined and one can further get
$Q$ from equation (\ref{xdotq}) or (\ref{omdotq}) and the spin of the
companion using $|\mbf{L}|$, $\theta_J$ and $\theta$ (see Fig.~9). This allows
to determine the moments of inertia $I_3^*$ and $I_1^*$ of the companion,
which are related with its internal structure.

For PSR B1259--63 the changes of $i$ and $\omega$ are typically of the order
of one second of arc per orbit (see Fig.~5). Since the orbital period
is 3.4 years, the quadratic-in-time changes of these angles will be absolutely
negligible for the next few decades. 

For PSR J0045--7319 the situation is different. The changes in $x$ and
$\omega$ are two orders of magnitude larger than for PSR B1259--63.  Fig.~10
shows the result of fitting for simulated pulse arrival times for PSR
J0045--7319.  The first figure (10a) gives the pre-fit residuals for a BT
model that leads to a good fit for the first few orbits. After four years the
model is off by about 40 ms. Fitting for the whole time span of 1500 days
using the BT model one finds the post-fit residuals given in the second figure
(10b). Most of the deviations of Fig.~10a are absorbed in the spin parameters
by changing them according to
\be
   \Delta     P = -2.1\times 10^{-10} \: {\rm s}\,, \qquad 
   \Delta\dot P = -3.3\times 10^{-18}\,.
\ee
The residuals are in the order of the present measurement precision. For
future observations the BT model (and therefore the BT++ and DD+ models) will
fail to explain the observations and one has to fit for higher derivatives in
$x$ and $\omega$. The result of such a fit is presented in the last figure
(10c).

\begin{figure} 
\vspace*{-0.5cm}
\psfig{figure=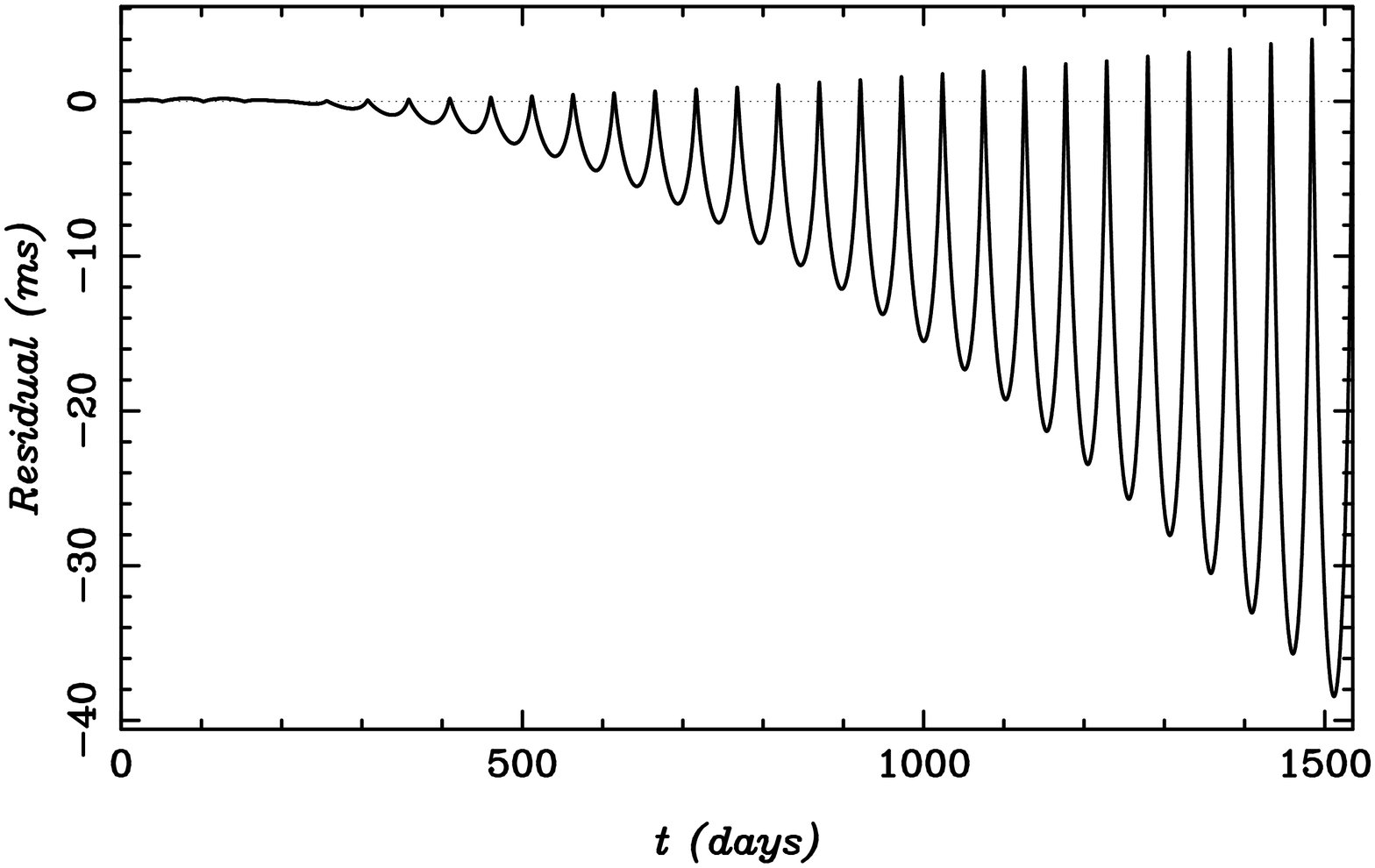,width=8.5cm,angle=-90}
\vspace{-0.5cm}
\psfig{figure=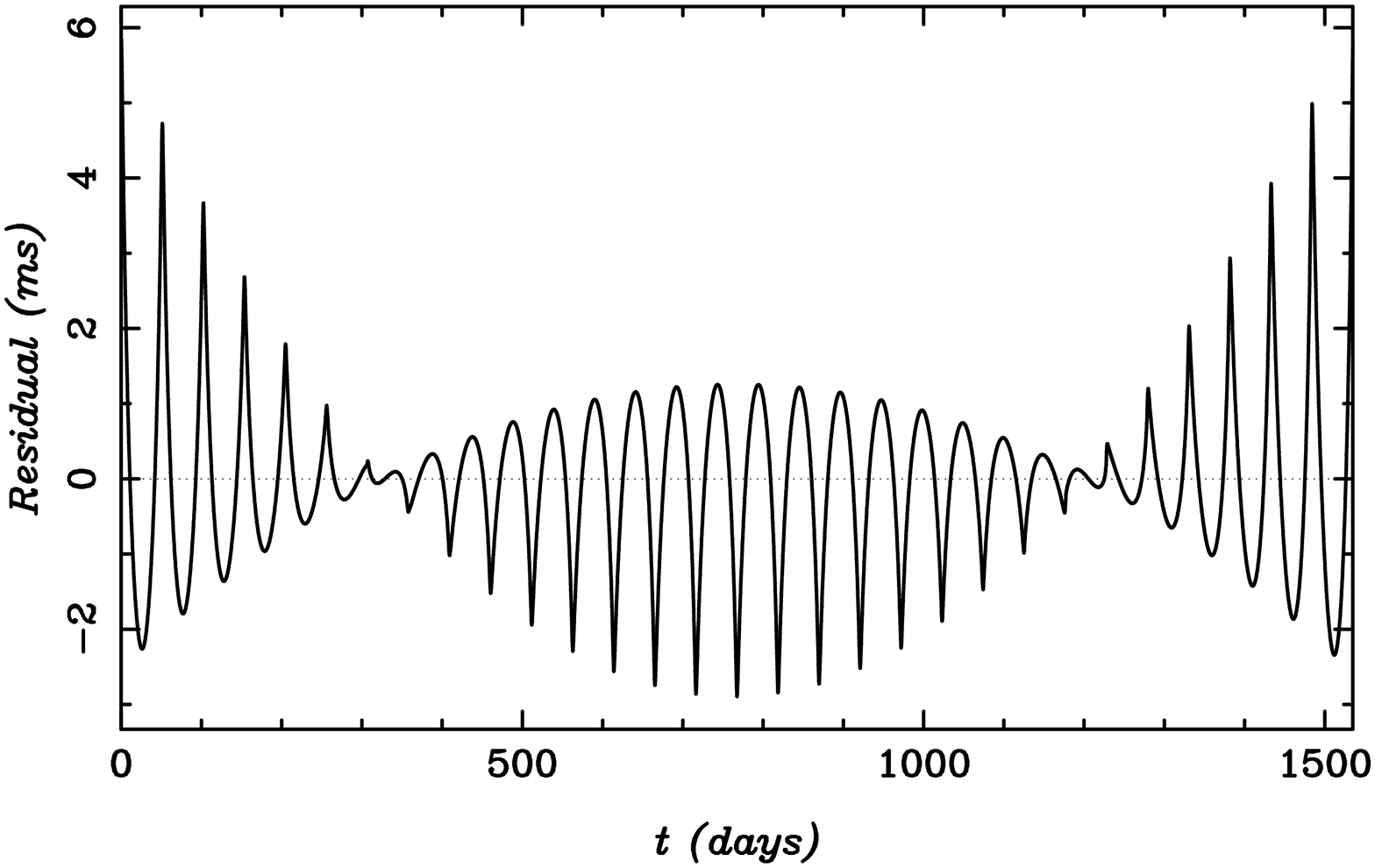,width=8.5cm,angle=-90}
\vspace{-0.5cm}
\psfig{figure=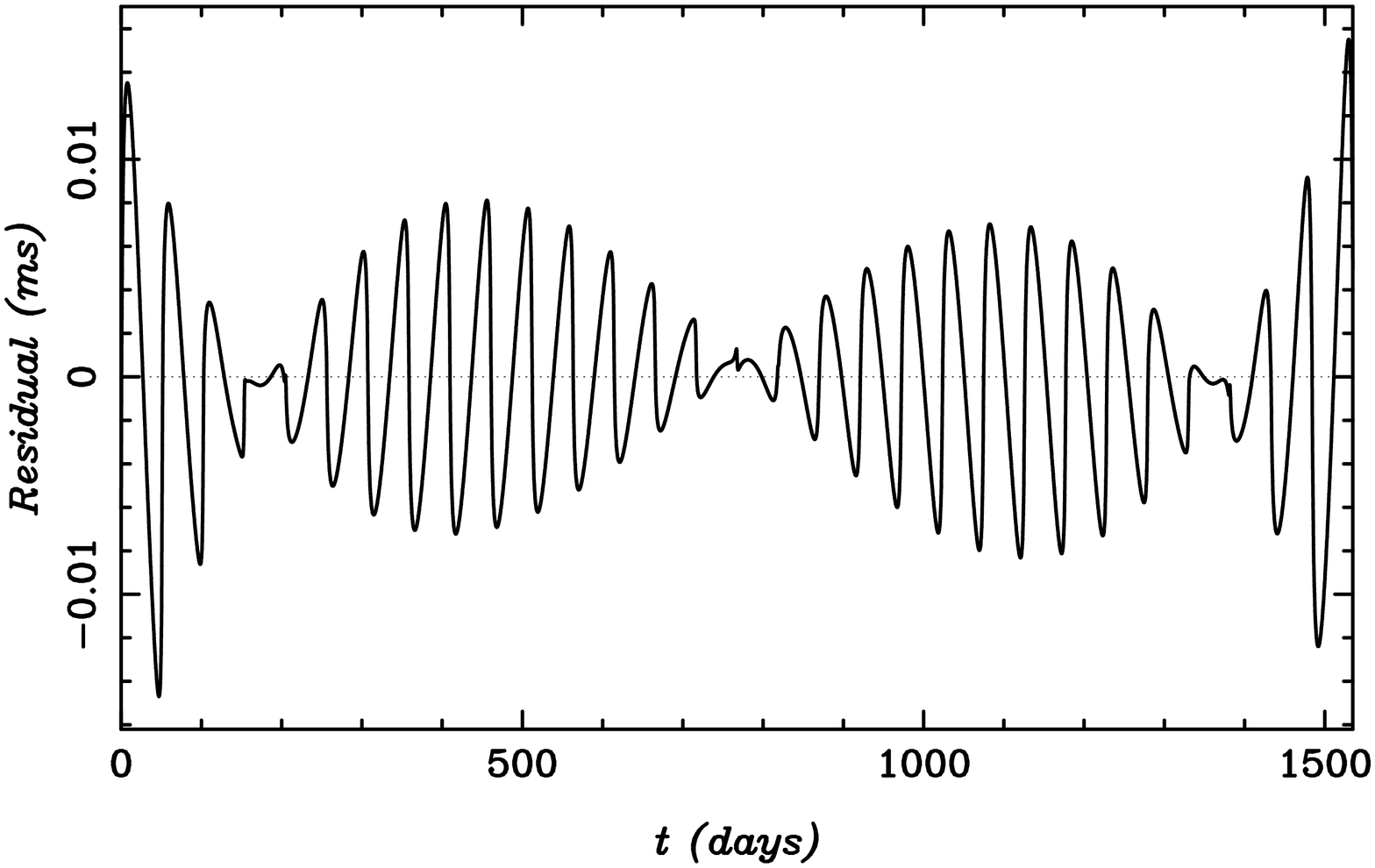,width=8.5cm,angle=-90}
\caption{Simulated pulse arrival times for PSR J0045--7319 over a time span of
	30 orbits ($\sim$ 4.2 years). {\it Top:} pre-fit residuals for a BT
        model which leads to a good fit for the first few orbits. {\it
        Middle:} post-fit residuals after fitting for spin and orbital
        parameters, including $\dot x$ and $\dot\omega$. {\it Bottom:}
        post-fit residuals after fitting for spin and orbital parameters,
        including $\dot x$ and $\dot\omega$, and second derivatives, $\ddot x$
        and $\ddot\omega$. I used the values of Table~1 ($i=44\degr$),
        $\theta=30\degr$, $\Phi_0=80\degr$, $k=0.01$, and
        $\hat\Omega_*=0.5$. $|\mbf{S}|/|\mbf{L}| \approx 0.1$ and thus
        $\theta_J\approx 3\degr$.}
\end{figure}

Finally, it should be mentioned that fitting for a $\dot P_b$ instead of $\ddot
x$ and $\ddot\omega$ improves the residuals only marginally and gives a $\dot
P_b$ which is two orders of magnitudes smaller than the one observed in this
system. Therefore the non-linear drifts of $x$ and $\omega$ cannot explain the
observed $\dot P_b$.


\section{PSR J0045--7319 --- Evidence for a neutron-star birth kick?}

The orbital precession of the binary pulsar PSR J0045--7319 was seen as a
direct evidence that the neutron star of this system received a kick of at
least 100 km/s at the moment of birth (Kaspi et al.\ 1996). Since the
theoretical analysis in this paper is based on the calculations of Smarr and
Blandford (1976) and Lai et al.\ (1995) an incorrect formula for $\dot\omega$,
the precession of the longitude of periastron, was used (equation (1) in Kaspi
et al.\ 1996). Therefore their limits on the angle between the spin axis of
the B star and the orbital angular momentum, $25\degr<\theta<41\degr$, are
incorrect and one has to reinvestigate the question whether this binary star
system provides an evidence for a neutron-star birth kick.

Using the values of Table~1 in Kaspi et al.\ (1996) one finds for PSR
J0047--7319
\be
   \frac{\dot\omega_0 x_0}{\dot x_0} = 1.80 \pm 0.05\,.
\ee
$\dot\omega$ was corrected for the general relativistic contribution of
$0.004\degr$/yr.  The uncertainty in the masses leads to
\be
   i=44\degr\pm5\degr \qquad ({\rm or} \quad i=136\degr\pm5\degr) \,.
\ee
Equation (\ref{ratio}) restricts the values of $\Phi_0$ and $\theta$ as shown
in Fig.~11 and one finds as a lower limit on the inclination of the B star
with respect to the orbital plane
\be
   \theta > 20\degr\,.
\ee
This limit is only slightly smaller than the one given in Kaspi et al.\ (1996)
and so does not change their major conclusion, i.e.\ that the binary system
PSR J0045--7319 provides direct evidence for a neutron-star birth kick of at
least 100 km/s. However, their conclusion that $\dot\omega_0>0$ implies
$\theta<55\degr$ is incorrect. In principle one can have $\theta$ up to
$70\degr$, although $\theta$ being close to $70\degr$ requires an unphysically
high apsidal motion constant $k$ for the B star. Numerical simulations show
that $\theta>65\degr$ is excluded by the fact that present timing observations
are still in agreement with a simple $\dot x$-$\dot\omega$ model. Similar
arguments apply for the case $\theta>90\degr$.

\begin{figure} 
\psfig{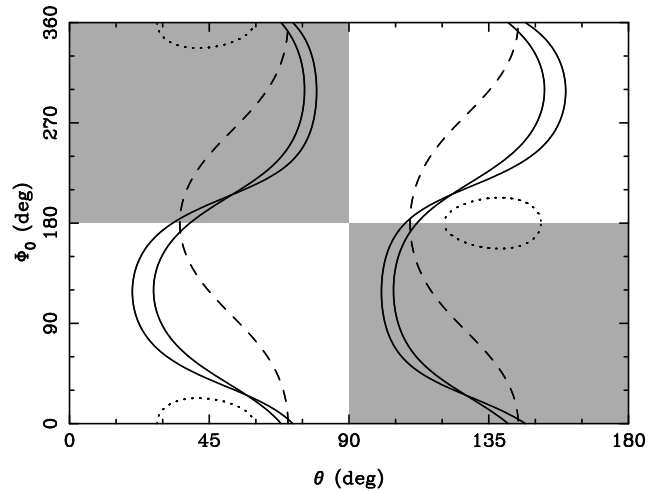}
\caption{Observational constraints on the geometry of the PSR J0045--7319 
binary system. I used $i=44\degr\pm5\degr$ (for $i=136\degr\pm5\degr$ one has
to replace $\Phi_0$ by $\Phi_0-180\degr$). The grey areas are excluded because
$\dot x>0$. The region between the two dashed curves is excluded because
$\dot\omega_0>0$. (To plot the dashed curves I used $i=44\degr$). The solid
curves border the narrow region of possible values of $\theta$ and
$\Phi_0$. The width of these regions is dominated by the error in $i$. The
regions within the dotted curves are ruled out since the proper rotation of
the B star (projected value: 113+10 km/s) does not exceed the break-up
velocity ($\sim$420 km/s), see Lai et al.\ 1995.  These regions do not lead to
any further constraints on $\theta$ and $\Phi$.}
\end{figure}


\section{Conclusions}

In this paper I have presented a timing formula for main-sequence star binary
pulsars which takes into account most of the periodic variations along the
orbit caused by the anisotropic nature of the $1/r^3$ potential of the
spin-induced quadrupole moment of the companion star. The new timing formula
contains the same number of parameters as the Blandford-Teukolsky timing
formula. I have shown by numerical simulations, that the new timing formula
leads to much better results in case of the long-orbital period binary pulsar
PSR B1259--63 then the Blandford-Teukolsky or the Damour-Deruelle timing
formula. Only very close to periastron the new timing formula shows deviations
are slightly greater than the typical measurement error in the time-of-arrival
of the pulsar signals. But so far there are no timing observations of PSR
B1259--63 close to periastron, due to the eclipse of the pulsar caused by the
circumstellar material. For PSR J0045--7319 these periodic variations are of
the order of the measurement precision.

I have given quadratic-in-time extensions of the timing formula which
account for long-term secular changes in the orientation of the binary-pulsar
orbit. In particular for the binary pulsar PSR J0045--7319 these extensions
might be important in the next few years of timing observations depending on
the, so far unknown, orientation of the B star spin and the total angular
momentum of the binary system. I have concluded that the measurement
of these long-term secular effects has the potential to probe the internal
structure of the companion.

Finally I have reinvestigated the classical spin-orbit precession of the
binary pulsar PSR J0045--7319 since the theoretical analysis of this binary
system given in Lai et al.\ (1995) and Kaspi et al.\ (1996) is based on an
incorrect expression for the precession of the longitude of periastron. I have
found $20\degr$ as a lower limit for the inclination of the B star with
respect to the orbital plane which does not change the conclusions concerning
the neutron-star birth kick in this system given in Kaspi et al.\ (1996).


\section*{Acknowledgments}

I thank Peter M\"uller for many stimulating discussions and Simon Johnston for
carefully reading the manuscript.


\end{document}